\documentclass{aa}
\usepackage[varg]{txfonts}
\usepackage{graphicx}
\usepackage{lscape}
\usepackage{multicol}
\usepackage{xcolor}
\usepackage{longtable}

\defcitealias{gon2015}{Paper~I}
\defcitealias{dor2016a}{Paper~II}
\defcitealias{dor2016b}{Paper~III}
\defcitealias{tab2018}{Paper~IV}

\begin{document}

\title{An atlas of cool supergiants from the Magellanic Clouds and typical interlopers\footnote{Table D is only available in electronic form
at the CDS, together with the atlas of spectra.}} 
\subtitle{A guide for the classification of luminous red stars}

\titlerunning{Atlas of cool supergiants in the Magellanic Clouds}

\author{Ricardo Dorda\inst{\ref{inst1}}
\and Ignacio Negueruela\inst{\ref{inst1}}
\and Carlos Gonz\'alez-Fern\'andez\inst{\ref{inst2}}
\and Amparo Marco\inst{\ref{inst1}}}
\institute{Departamento de F\'{\i}sica, Ingenier\'{\i}a de Sistemas y Teor\'{\i}a de la Se\~nal, Universidad de Alicante, Carretera de San Vicente s/n, San Vicente del Raspeig E03690, Alicante, Spain\label{inst1}
\and Institute of Astronomy, University of Cambridge, Madingley Road, Cambridge CB3 0HA, United Kingdom\label{inst2}
}

\abstract{We present an atlas composed of more than $1\,500$ spectra of late-type stars (spectral types from G to M) observed simultaneously in the optical and calcium triplet spectral ranges. These spectra were obtained as part of a survey to search for cool supergiants in the Magellanic Clouds and were taken over four epochs. We provide the spectral and luminosity classification for each spectrum  (71\% are supergiants, 13\% are giants or luminous giants, 4\% are carbon or S~stars, and the remaining 12\% are foreground stars of lesser luminosities). We also provide a detailed guide for the spectral classification of luminous late-type stars, the result of the extensive classification work done for the atlas. Although this guide is based on classical criteria, we have put them together and re-elaborated them for modern  CCD-spectra as these criteria were scattered among many different works and mainly conceived for use with photographic plate spectra. The result is a systematic, well-tested process for identifying and classifying luminous late-type stars, illustrated with CCD spectra of standard stars and the classifications of our own catalogue.}


\maketitle

\section{Introduction}

Cool supergiants (CSGs) are evolved stars of moderately high mass (between $\sim10$ and $\sim40\:$M$_{\odot}$), characterized by their very high luminosities \citep[$\log(L/L_{\sun})\sim4.5$\,--\,5.8;][]{hum1979} and late spectral types (G, K, and M). There are not many of these stars, due to the intrinsic scarcity of high-mass stars, but also because the cool supergiant phase is just a small fraction of their short total lifespans \citep[between $\sim8$ and $\sim25\:$Ma;][]{eks2013}. Until recently, only a few hundred of these stars were known: between 100 and 200 in the Galaxy \citep[e.g.\ ][]{hum1978,lev2005,fig2006,dav2007,dav2008,cla2009a,neg2011,neg2012}, a few hundred in the Magellanic Clouds \citep{hum1979b,mas2003b,neu2012}, and $\sim300$ in M33 \citep{dro2012}.

To study the properties of CSGs as a population, our group performed an ambitious spectroscopic survey of the Magellanic Clouds (MCs). We obtained almost $1\,500$ spectra along four epochs. All our targets were observed simultaneously in the optical and calcium triplet (CaT) spectral ranges. In \citet[hereafter Paper~I]{gon2015} we presented the target selection and observations, the radial velocity calculations, and the spectral and luminosity classification that were performed. From the combination of the last two, we discussed the nature of the targets, as well as their membership in the Clouds. The resulting catalogue contains the largest sample of CSGs uniformly observed to date (over 500 unique objects), but also includes a similar number (almost 450 spectra\footnote{We found 333 interlopers in the SMC fields and 109 in the LMC fields. These interlopers represent 31\% (SMC) and 26\% (LMC) of the spectra observed towards each galaxy.}) of typical interlopers, stars whose photometric properties make them undistinguishable from CSGs through photometric criteria. In \citetalias{gon2015}, we also analysed the selection efficiency of our photometric criteria and the completeness of the CSG sample. We studied the position of CSGs in colour-magnitude diagrams based on \textit{WISE} mid-infrared data with the aim of improving selection criteria. In addition, we compared our sample and the efficiency of our criteria at detecting CSGs with those in previous works, based on either spectroscopic or photometric criteria.

The CSG sample found in \citetalias{gon2015} was studied in detail in successive papers. \citet[Paper~II]{dor2016a} analysed statistically the spectral properties of the CSGs in the sample and characterized them as a population. For this analysis, we studied the behaviour of the main spectral features in the CaT range, by comparison with synthetic models, and its relation with the spectral types (SpTs) and luminosity classes (LCs) derived in \citetalias{gon2015} from the optical-range spectra. We also characterized the SpT distribution for CSGs in each Galaxy, and the extent of spectral variability in those CSGs observed on more than one epoch.

Later, \citet[Paper~III]{dor2016b} presented an automated method for the identification of CSGs based on principal component analysis (PCA). Using measurements of the main spectral features in the CaT region for the whole sample (CSGs and non-supergiants), we calculated the principal components for each star. Relating their behaviour to the LCs obtained in \citetalias{gon2015}, we obtained a method to split CSGs from non-supergiants in the parameter space of principal components. The result is a highly efficient automated method that separates CSGs from typical interlopers, with the advantage over other methods used in the literature (such as the use of CaT equivalent widths) that it keeps the number of contaminants close to zero.

Finally, \citet[Paper~IV]{tab2018} analysed the CSGs in the sample through spectral synthesis (in the CaT range), deriving their physical properties. We compared the SpTs and LCs assigned in \citetalias{gon2015} with the physical parameters obtained through this analysis. This allowed the calculation of the temperature scale, i.e.\ the relation between SpT and effective temperature for the CSGs in both clouds, and the metallicity and effective temperature distributions for these stars. We also included an analysis of the effect of spectral variability on the effective temperature of these stars.

This is the final paper in the series. Here we present an atlas containing all the spectra observed in the direction to the MCs, and make them available to the whole community. There are not many spectra of CSGs in existing spectral atlases \citep[e.g.][]{val2004,sbl2006} nor in papers on spectral classification \citep{car1997,kir1991}. Therefore, we expect this work to be useful for a wide audience because we  provide a large number of spectra,  large enough to perform statistical analyses. In addition, our catalogue contains multi-epoch observations for $\sim200$~CSGs, which can be useful for any study of spectral variability in CSGs. Finally, this atlas also contains all the spectra of the non-CSG interlopers that we observed ($\sim300$ spectra between carbon stars, luminous giants, and less luminous stars). We expect these objects to be useful for a broad range of astrophysical topics. In addition, these stars can be used for classification purposes as they represent the typical interlopers selected as candidates to CSGs through photometric criteria. Furthermore, we have cross-matched  our catalogue with the recent \textit{Gaia} DR2 database \citep{brown18}, which provides for the first time accurate positions and proper motions for all the stars in the catalogue, as well as radial velocities (RVs) for the vast majority of them.

We also include in this paper a very detailed guide to the classification of CSGs in the optical range, also including  multiple figures to illustrate the classification process. A guide of this kind has been in strong demand within the community interested in luminous cool stars. This method is the very same that we used in \citetalias{gon2015}, but there we provided only a brief description of the method, simply aimed at showing that we had used classic classification criteria. The methodology described here is not fully original, but rather a synthesis of classic criteria which are scattered among many different works \citep{mor1943,fit1951,kee1976,kee1977,kee1980,mor1981,tur1985,kee1987} and were developed for use with photographic plates. We searched through the literature on spectral classification in the optical range and put together  all the criteria in a homogeneous way. Moreover, we tested the reliability and usefulness of the criteria when applied to a large sample of modern CCD-spectra (both those of standard stars and our own observations). Since our optical spectra were observed at different resolving powers, we explore the usefulness of criteria in each case, providing a wide perspective on the classification of luminous cool stars. The result is a systematic, well-tested process for identifying and classifying luminous late-type stars, illustrated with CCD spectra of standard stars and the classifications of our own catalogue.

This paper is organized as follows. In Section~\ref{data} we explain the observations, data processing, and the composition of the atlas. In Section~\ref{spec_class} we give a detailed explanation of the classification process used that will be useful for those interested in the identification and classification of CSGs.

\section{Description of the atlas}
\label{data}

The observational strategy and reduction process are explained in detail in \citetalias{gon2015}. In this paper we do not revisit  the target selection criteria as they hold no interest for the present atlas. However, we provide an overview of the observations as it can be useful to understand the composition of the present atlas.

\subsection{Observations}
\label{obs}

The instrument used for the observations was the fibre-fed dual-beam AAOmega spectrograph on the 3.9~m Anglo-Australian Telescope (AAT). We did four different observation campaigns (two for the SMC alone, one  for the LMC alone,  and one for both MCs together). During our campaigns we repeatedly observed about a hundred well-known red supergiants (RSGs) from each MC. We also observed several hundred CSG  candidates  in one campaign only for each MC (2012 for the SMC and 2013 for the LMC). All the non-supergiants included in our atlas come from these epochs. 

AAOmega can observe up to 400 targets simultaneously in the optical and the nearest infrared ranges (in the region of the infrared CaT). For the infrared range we used the same grating, 1700D, in all epochs. It provides a range that is  500~\AA{} in width with a nominal resolving power ($\lambda/\delta\lambda$) of $11\,000$ at the wavelengths considered. We centred the blaze on $8600$~\AA{} for the 2010 observations, but on $8700$~\AA{} in all other epochs. The optical range observations were done with two different gratings in different campaigns: 580V covers a wide range (about 2100~\AA{}) with $\lambda/\delta\lambda\sim1\,300$, while the range covered by 1500V is narrower but with a higher resolution power ($\lambda/\delta\lambda3\,700$). The observations are summarized in Table~\ref{obsconf}.

\begin{table*}
\caption{Summary of the observations}
\label{obsconf}
\centering
\begin{tabular}{c | c c c c | c c c c}
\hline\hline
\noalign{\smallskip}
&\multicolumn{4}{c|}{Blue arm}&\multicolumn{4}{c}{Red arm}\\
Year&Grating&$\lambda_{{\rm cen}}$ (\AA{})&Range (\AA{})&$\lambda/\delta\lambda$&Grating&$\lambda_{{\rm cen}}$ (\AA{})&Range (\AA{})&$\lambda/\delta\lambda$\\
\noalign{\smallskip}
\hline
\noalign{\smallskip}
2010&580V&4500&$\sim2\,100$&$\sim1\,300$&1700D&8600&$\sim500$&$\sim10\,000$\\
2011&1500V&4400&$\sim800$&$\sim3\,700$&1700D&8700&$\sim500$&$\sim10\,000$\\
2012&1500V&5200&$\sim800$&$\sim3\,700$&1700D&8700&$\sim500$&$\sim10\,000$\\
2012&580V&4800&$\sim2\,100$&$\sim1\,300$&1700D&8700&$\sim500$&$\sim10\,000$\\
2013&580V&4800&$\sim2\,100$&$\sim1\,300$&1700D&8700&$\sim500$&$\sim10\,000$\\
\noalign{\smallskip}
\hline
\end{tabular}
\end{table*}

Reduction of the data was done through the standard automatic reduction pipeline {\tt 2dfdr} as provided by the AAT at the time. Wavelength calibration was performed with arc lamp spectra (He+CuAr+FeAr+ThAr+CuNe) immediately before observing each field. The resulting calibrations always have the root mean square of the error  under $<0.1$ pixels.

Finally, is important to indicate that none of our spectra is flux-calibrated or normalized. The flux calibration was not done because it is not an easy task in the case of a multi-object spectrograph, and it was not necessary for our scientific objectives. For the classification we only used local normalizations as the presence of TiO bands eroding the continuum in many spectra makes  their normalization complicated. Thus, we think that anyone interested in these spectra would need their particular normalization, depending on the spectral ranges of interest.

\subsection{Radial velocities}

We provide the spectra as originally observed, without any radial velocity (RV) correction, to allow any recalculation or study of the radial velocity distribution. However, we include the RVs that we calculated in \citetalias{gon2015} for all the spectra in our catalogue. Therefore, any spectrum can be easily shifted to the rest frame.

To calculate RVs we followed the procedure outlined in \citet{kop11} applied to the spectra from the red arm. This method uses sky emission to refine possible systematics remaining in the wavelength calibration, so that all the spectra are in a common system that  can  be anchored by using velocity standards. Then it compares the observed spectra with a battery of models using a Bayesian framework. Through this method it is possible to marginalize over any parameter in which we are not interested (in this case, for example, the continuum normalization and the reference model), removing it from the analysis while at the same time taking it into account when deriving uncertainties. In our case, we marginalize over the continuum normalization, since continuum determination is almost impossible at these resolutions for late-type stars, and over the stellar model, so that the derived velocities are not model dependent.

\begin{figure}[ht!]
   \centering
   \includegraphics[trim=0.3cm 0.2cm 1.2cm 1.2cm,clip,width=7.5cm]{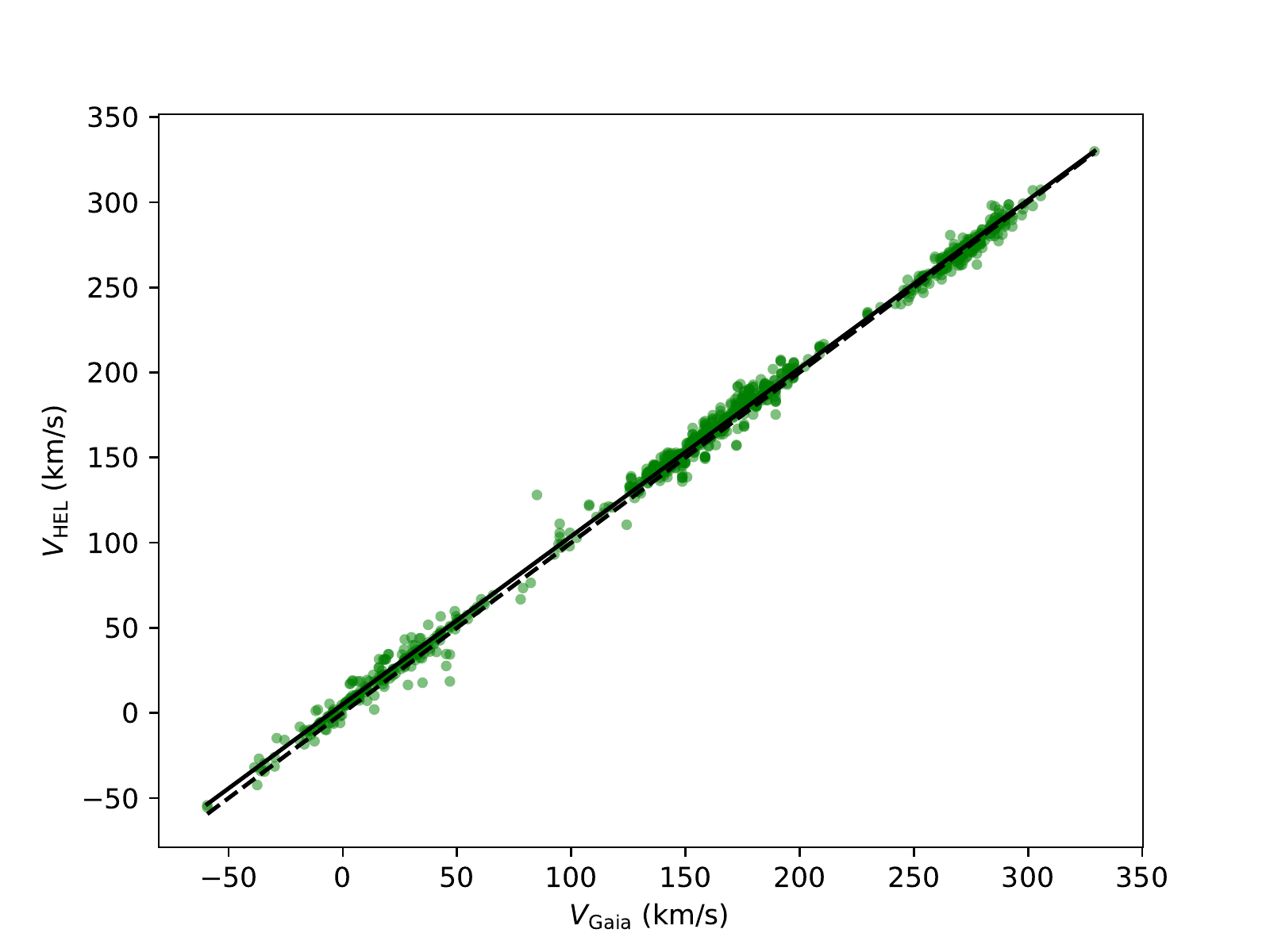}
   \includegraphics[trim=0.3cm 0.2cm 1.2cm 1.2cm,clip,width=7.5cm]{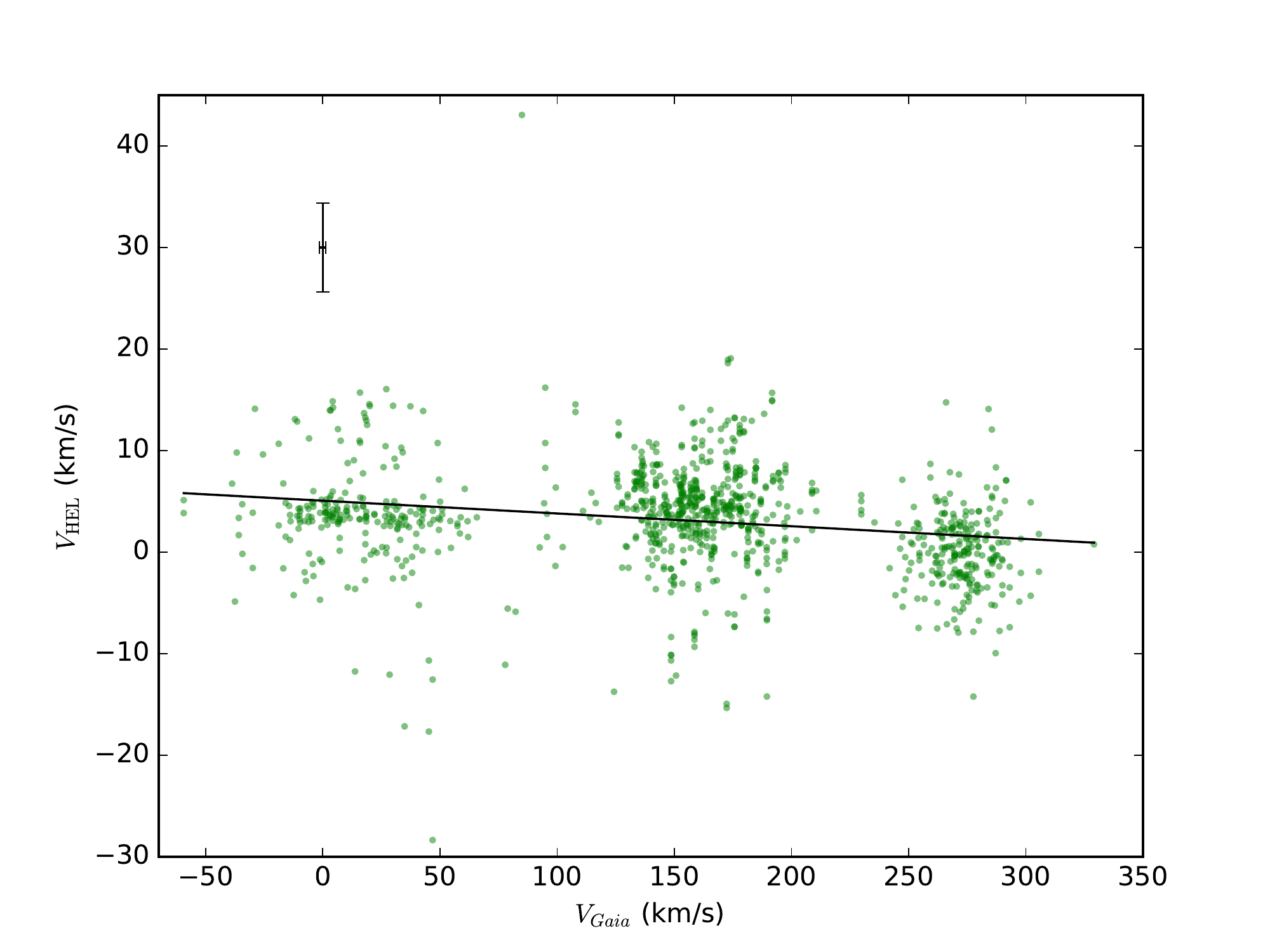}
   \caption{Comparison between radial velocities from \textit{Gaia} ($V_{Gaia}$) for the targets in this catalogue and those obtained through our spectra ($V_{\rm HEL}$). The individual measurements of stars with several measurements in our catalogue are shown as separate points.
   {\bf Left (\ref{Vrad}a):} $V_{Gaia}$ vs. $V_{\rm HEL}$. The dashed line indicate the ratio 1 to 1, while the solid line is the corresponding to the linear regression ($[m\pm e_m]x + [n\pm e_n] = [0.9870\pm0.0018]x + [5.1\pm0.3]$). The error bars are too small to be represented.
   {\bf Right (\ref{Vrad}b):} $V_{Gaia}$ vs. $V_{Gaia}-V_{\rm HEL}$. The solid line is  corresponds to the linear regression indicated in the left panel. The black cross indicates the average error on the data on each axis.
   }
   \label{Vrad}
\end{figure}

To check the general quality of our RVs calculated in \citetalias{gon2015}, we have compared them with those given in \textit{Gaia} DR2 for our targets. We found that 71\% of our spectra present a RV in \textit{Gaia} (among CSGs the fraction is slightly higher, 76\%). Figure~\ref{Vrad} shows a comparison of our data to those of DR2. Except for a few outliers the correlation is excellent. We performed a linear regression (through the least-squares method) and obtained $[m\pm e_m]x + [n\pm e_n] = [0.9870\pm0.0018]x + [5.1\pm0.3]$. Although this fit seems to indicate a systematic offset by $5.1\pm0.3\:\mathrm{km}\,\mathrm{s}^{-1}$ between our RVs and those of \textit{Gaia}, Fig.~\ref{Vrad} shows that it decreases for higher RVs, becoming compatible with our average uncertainty for RV (see Sect.~\ref{uncert}) for stars with RVs typical of the MCs (RV$>90\:\mathrm{km}\,\mathrm{s}^{-1}$). The average (and standard deviation) of $\Delta V=V_{Gaia}-V_{\rm HEL}$ is $0\pm4\:\mathrm{km}\,\mathrm{s}^{-1}$ for the CSGs in the LMC and $4\pm5\:\mathrm{km}\,\mathrm{s}^{-1}$ for those in the SMC. Stars with RV$<90\:\mathrm{km}\,\mathrm{s}^{-1}$ are foreground to the clouds, as evidenced by parallaxes generally indicating distances smaller than 2~kpc, with the single exception of the K4\,Ia supergiant SMC091 = PMMR~10, one of the brightest RSGs in the SMC, whose peculiar velocity was noted in \citetalias{gon2015}. Its astrometric parameters fully confirm that this is an SMC member.

The apparent offset is most likely caused by some degree of scatter among foreground stars. The stars showing higher differences between our values and those of \textit{Gaia} are mostly foreground G dwarfs. Two of them, SMC165 and SMC202, were already identified as RV variables in \citetalias{gon2015}, and so we believe that the scatter is real and caused by binarity. There is, however, an SMC supergiant for which the \textit{Gaia} RV is very different from our value. This is YSG010 = LHA 115-S 7, a G-type supergiant with emission lines. Its blue spectrum shows the presence of a B-type star, and it was marked as a candidate interacting binary in \citetalias{gon2015}. The large difference in RV between our data and DR2 is further evidence in this sense.

As an example of the scientific potential of cross-matching our data with \textit{Gaia} DR2, we highlight the case of 168290. The RV of this star is about 305~$\mathrm{km}\,\mathrm{s}^{-1}$ in both our data and DR2. This value is high when compared with the average value for the CSGs from the LMC \citepalias[$271\pm15$;][]{gon2015}. Moreover, its proper motions from \textit{Gaia} (pmRA$=1.53\pm0.05$ pmDEC$=0.83\pm0.06$) are significantly different from those typical of the LMC. In contrast, its spectra  unequivocally show an LC~I star, while its magnitudes are typical of an LMC supergiant. If we calculate its peculiar proper motions with respect to the bulk of CSGs in the LMC, we find that it points towards the edge of the galaxy. Taken all together,  these factors make 168290 a strong candidate runaway massive star.

\subsection{Composition of the atlas}
\label{compos}

In \citetalias{gon2015} we presented a catalogue with information on individual stars. Some stars had been observed on multiple occasions during a given observing run (typically once per night). For those objects, we provided average values of RV, spectral type (SpT), and luminosity class (LC) calculated through a mean of the individual spectra weighted by their signal-to-noise ratio (S/N). The spectra of stars observed on different epochs were checked for spectral variability. When not tagged as spectrally variable, the values from different epochs were also averaged.

In contrast to this procedure, in this atlas we present all the individual spectra together with their individual RVs, SpTs, and LCs. To relate them to the catalogue already published in \citetalias{gon2015}, we provide the identification in that catalogue for each target. Many of the targets have been observed more than once; thus there are many spectra with the same catalogue name. In consequence, for the unequivocal identification of each target, we provide three datapoints: epoch, field, and fibre number in that field (see Table~\ref{tab_total}). We note that for each unique identification there are two spectra, one from the blue arm and one from the red arm, which were observed simultaneously.

This atlas includes almost $1\,500$ spectra (see Table~\ref{sum_MC} for a detailed description). Most of them (more than $1\,000$) correspond to the $\sim500$ unique CSGs from the MCs identified in the catalogue. The rest correspond to the interlopers found in the survey because they passed the cut of our photometric criteria. They therefore represent the kind of interlopers that a survey looking for CSGs will have to handle. Among them, there is a large sample of relatively luminous stars: almost 200 giants and luminous giants (13\% of the total atlas) and more than 60 carbon stars (4\%), mostly from the MCs. The remainder of the atlas corresponds to a slightly smaller number of less luminous stars that do not belong to the MCs (by LC and by RV, see \citetalias{gon2015}), but rather to the Galactic foreground (12\%).

\begin{table*}
\caption{Summary of the spectra observed and classified in our work. They have been split by field (galaxy), epoch, LC, and SpT.}
\label{sum_MC}
\centering
\begin{tabular}{c | c | c c c c c}
\hline\hline
\noalign{\smallskip}
Field&Spectral&\multicolumn{5}{c}{Number of}\\
(epoch)&Type&LC I&LC II\,--\,III&LC IV\,--\,V&C/S~stars&Total\\
\noalign{\smallskip}
\hline
\noalign{\smallskip}
                &G&81&0&0&--&81\\
SMC     &K&164&2&0&--&166\\
(2010)  &M&28&0&0&--&28\\
                &All&273&2&0&0&275\\
\noalign{\smallskip}
\hline
\noalign{\smallskip}
                &G&10&0&0&--&10\\
SMC             &K&90&1&0&--&91\\
(2011)  &M&4&0&0&--&4\\
                &All&104&1&0&0&105\\
\noalign{\smallskip}
\hline
\noalign{\smallskip}
                &G&135&33&96&--&164\\
SMC             &K&182&76&48&--&306\\
(2012)  &M&44&9&8&--&61\\
                &All&362&138&178&14&692\\
\noalign{\smallskip}
\hline
\noalign{\smallskip}
                &G&226&33&96&--&335\\
Totals  &K&436&79&48&--&563\\
for SMC &M&76&9&8& &93\\
                &All&739&141&178&14&1072\\
\noalign{\smallskip}
\hline
\noalign{\smallskip}
                &G&1&0&0&--&1\\
LMC             &K&28&0&0&--&28\\
(2010)  &M&55&0&0&--&55\\
                &All&84&0&0&0&84\\
\noalign{\smallskip}
\hline
\noalign{\smallskip}
                &G&6&2&0&--&8\\
LMC             &K&94&16&4&--&114\\
(2013)  &M&124&37&1&--&162\\
                &All&225&55&5&49&334\\
\noalign{\smallskip}
\hline
\noalign{\smallskip}
                &G&7&2&0&--&9\\     
Totals  &K&122&16&4&--&142\\
for LMC &M&179&37&1&--&217\\
                &All&309&55&5&49&418\\
\noalign{\smallskip}
\hline
\end{tabular}
\end{table*}

\subsection{Uncertainties}
\label{uncert}

All the targets were classified using their optical spectra following the process detailed in Section~\ref{spec_class}. The infrared spectra were used for other purposes \citepalias{dor2016a,dor2016b,tab2018}. In this atlas we only use the infrared spectra  to calculate radial velocities (RVs). There is some overlap between different observations in the same epoch. As we performed the classification for each of the spectra of these redundant targets independently, we can use them to test the internal coherence of our classification scheme. We obtained typical differences of one subtype in SpT and half a subclass in LC. For the RV, we obtained a typical dispersion of $ \sim1.0~\mathrm{km\,s^{-1}}$. In consequence, we assumed a $99\%$ confidence interval for our measurements of RV at $4~\mathrm{km\,s^{-1}}$. 

For the spectral classification of our CSGs we obtained $\pm1$ in SpT and $\pm0.5$ in LC. However, while the RV dispersion is applicable to all cases, the uncertainties on LC and especially on SpT can only be applied to MC supergiants. We cannot quantify the value of the uncertainties on SpT and LC for the interlopers because almost all the targets observed more than once in the same epoch are supergiants. Thus, the uncertainties derived for supergiants can be taken as a lower limit for the uncertainties of non-supergiants, but probably slightly higher uncertainties should be expected in stars with LC~III to LC~V because these stars were not the main objective of this work and we did not require the same level of quality  in their classification.

\section{Spectral and luminosity classifications}
\label{spec_class}

This classification guide is focused on luminous (LC I to III), cool stars (types G, K, and M). We include red giants and luminous giants (LCs III and II, respectively) because any useful classification criteria should be able to separate these stars from the CSGs. Although S~and C~stars present bolometric magnitudes similar to luminous giants, we do not include them in this guide because their spectra are very characteristic and easily distinguishable from oxygen-rich giants and cool supergiants. As we did not have any scientific interest in them, we simply tagged them as carbon-rich giants, but we did not analyse them in  detail. They are collectively termed `C/S stars' in Table~\ref{sum_MC}.

Classical classification criteria are defined in the spectral range covered by our blue-arm spectra  \citep{mor1943,fit1951,kee1976,kee1977,kee1980,mor1981,tur1985,kee1987}. These criteria were complemented with our own secondary indicators, whose variation with SpT and LC was derived through visual comparison of high-quality spectra of standard stars. These were taken from the Indo-US spectral library \citep{val2004} and the MILES star catalogue \citep{sbl2006}, and degraded to the spectral resolution corresponding to our observations. The complete list of standard stars used for the classification is shown in Table~\ref{standard}. All the figures in this section show the spectra of standard stars, rather than objects from our own catalogue, as these are the spectra used to build the classification criteria that were later applied to our sample. A sample of spectra from our own catalogue is shown in Appendix~\ref{spec_sample} illustrating the spectral sequence for supergiants of both MCs.

\citet{hum1979a} reported that the metallicity differences among the Milky Way (MW), the LMC, and the SMC do not change the behaviour of the atomic line ratios and other spectral features used in the spectral classification of CSGs. It is thus possible to use MW standard stars as a comparison, and the same criteria developed for CSGs in one galaxy are applicable to the others as long as they are based on line (or band) ratios and not line (or band) strengths. With our sample we can confirm that there are no noticeable differences in the ratios and spectral features used for classification between the CSGs from both MCs due to metallicity. In consequence, we adopted the same criteria for both MCs, using Galactic standard stars as comparison. To ease the use of this guide, in Table~\ref{lines} we provide a complete list of atomic lines and molecular bands used in the classification process.

\begin{figure*}
   \centering
   \includegraphics[trim=1.8cm 0.4cm 1.5cm 1.3cm,clip,width=12.5cm]{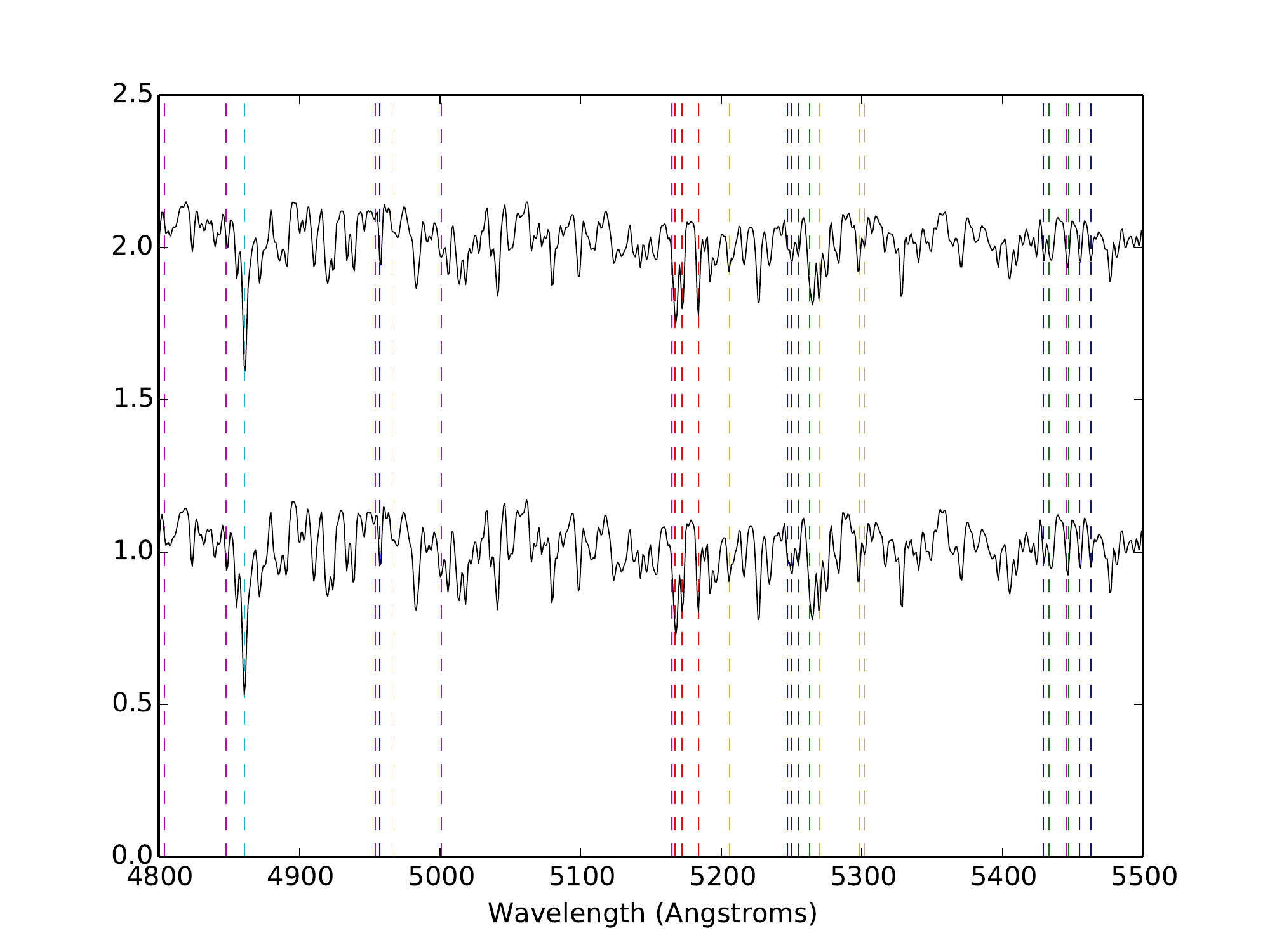}
   \caption{Example of the effect of luminosity in early-G stars. The stars are shown  from bottom to top: HD~74395 (G1\:Ib) and HD~84441 (G1\:II). Vertical lines indicate the position of spectral features useful for classification. Their colours represent the dominant chemical species in each feature: red for Mg\,{\sc{i}} (the Mg~Triplet), blue for Fe\,{\sc{i}}, green for Ti\,{\sc{i}} and Mn\,{\sc{i}}, yellow for other atomic lines (Ca\,{\sc{i}}, Cr\,{\sc{i}}, V\,{\sc{i}} and {\sc{ii}}, and blends), cyan for Balmer's H~lines, and magenta for the TiO bands (see text for  details).}
   \label{MgT_G}
\end{figure*}

\begin{figure*}
   \centering
   \includegraphics[trim=1.8cm 0.4cm 1.5cm 1.3cm,clip,width=12.5cm]{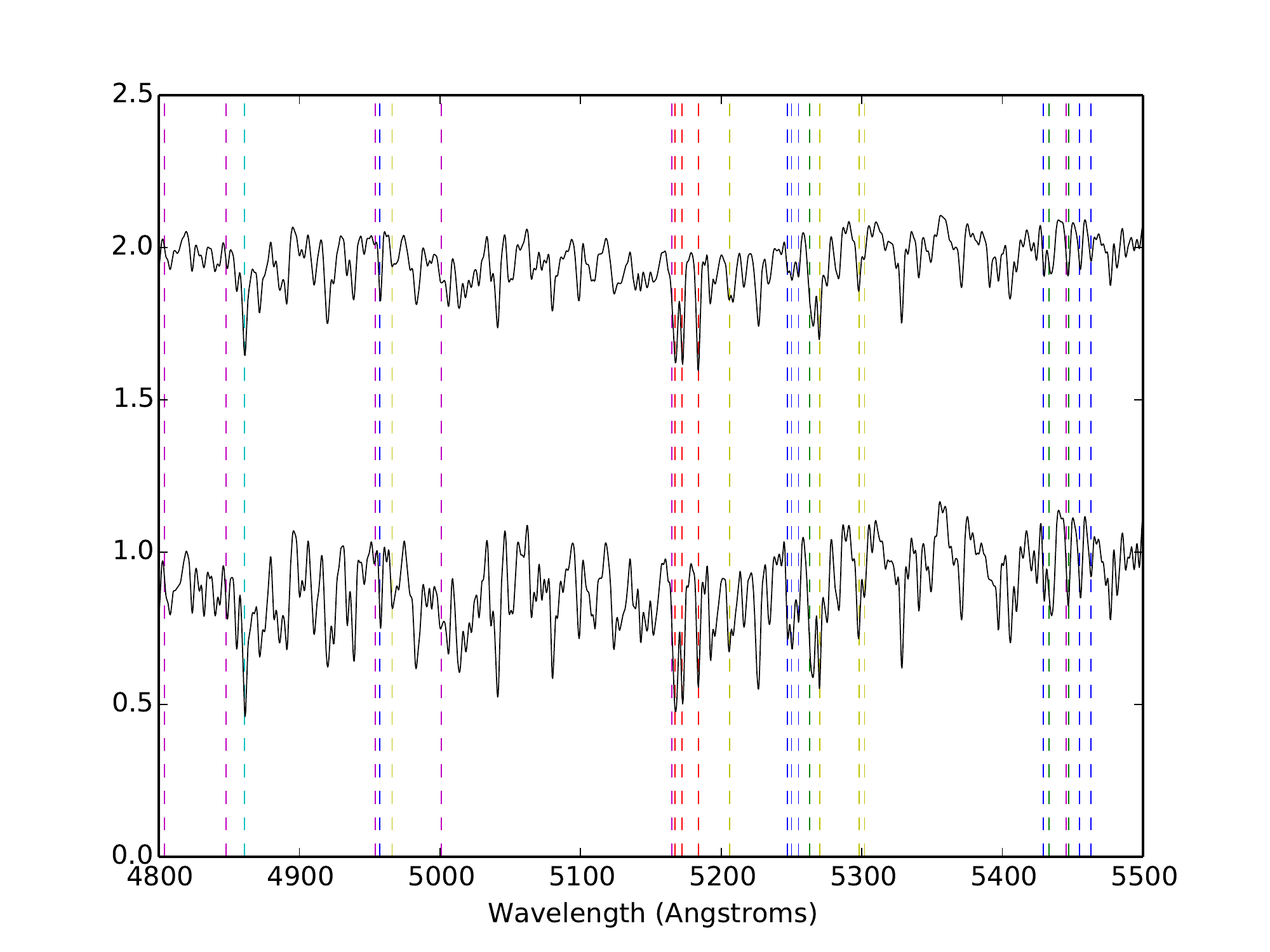}
   \caption{Example of the effect of luminosity in late-G stars. The stars are shown  from bottom to top: HD~48329 (G8\:Ib) and HD~5516 (G8-\:III). Dashed lines are as in Fig.~\ref{MgT_G}.}
   \label{MgT_G_2}
\end{figure*}

\begin{figure*}
   \centering
   \includegraphics[trim=1.8cm 0.4cm 1.5cm 1.3cm,clip,width=12.5cm]{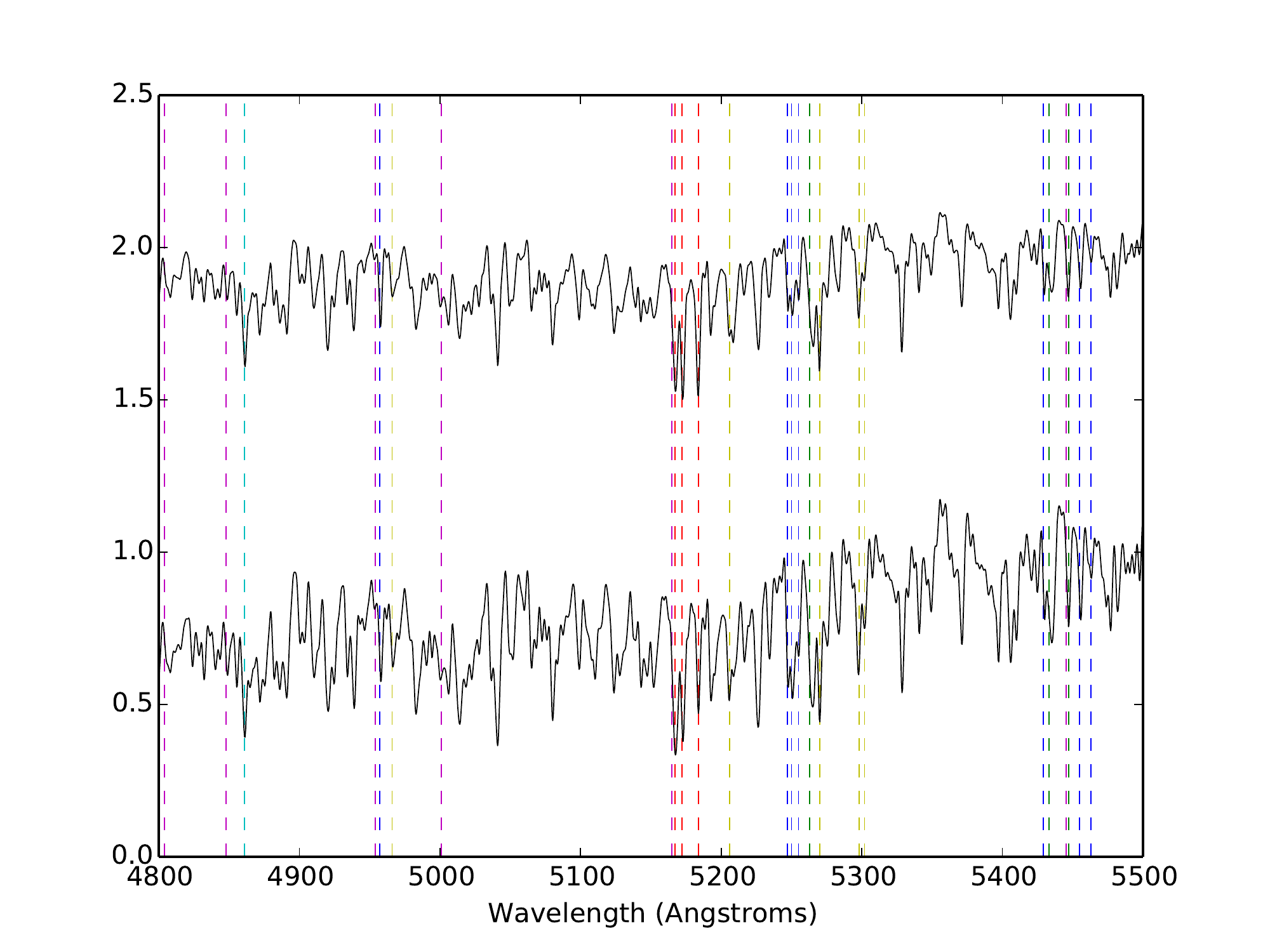}
   \caption{Example of the effect of luminosity in early-K stars. The stars are shown  from bottom to top:  HD~63302 (K1\:Ia\,--\,Iab) and HD~43232 (K1\:III). Dashed lines are as in Fig.~\ref{MgT_G}.}
   \label{MgT_K}
\end{figure*}

\begin{figure*}
   \centering
   \includegraphics[trim=1.8cm 0.4cm 1.5cm 1.3cm,clip,width=12.5cm]{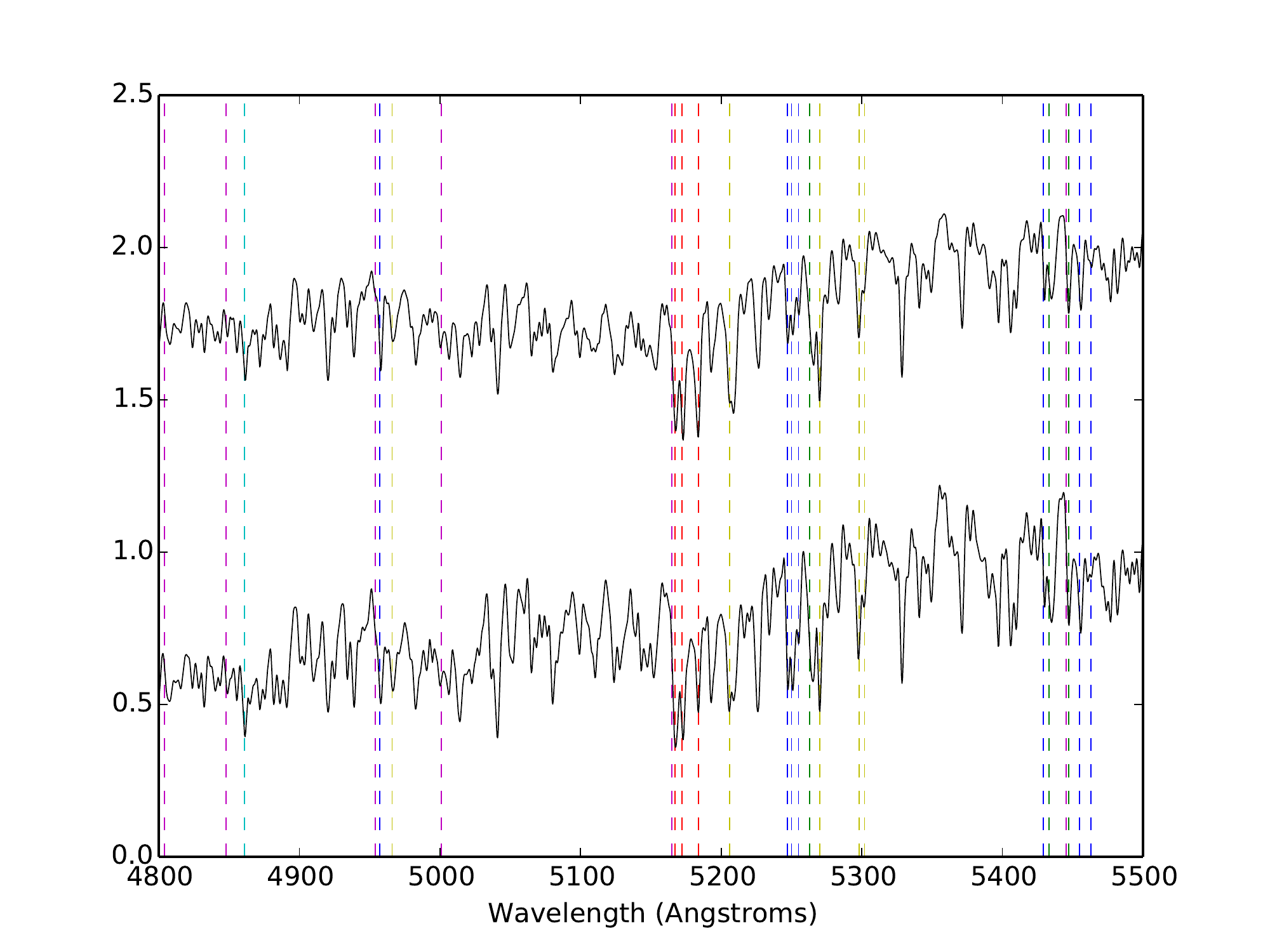}
   \caption{Example of the effect of luminosity in late-K and early-M stars. The stars are shown  from bottom to top: HD~44537 (K5\,--\,M1\:Iab\,--\,Ib) and HD~113996 (K5-\:III). Dashed lines are as in Fig.~\ref{MgT_G}.}
   \label{MgT_M}
\end{figure*}

\begin{figure*}
   \centering
   \includegraphics[trim=1.8cm 0.4cm 1.5cm 1.3cm,clip,width=12.5cm]{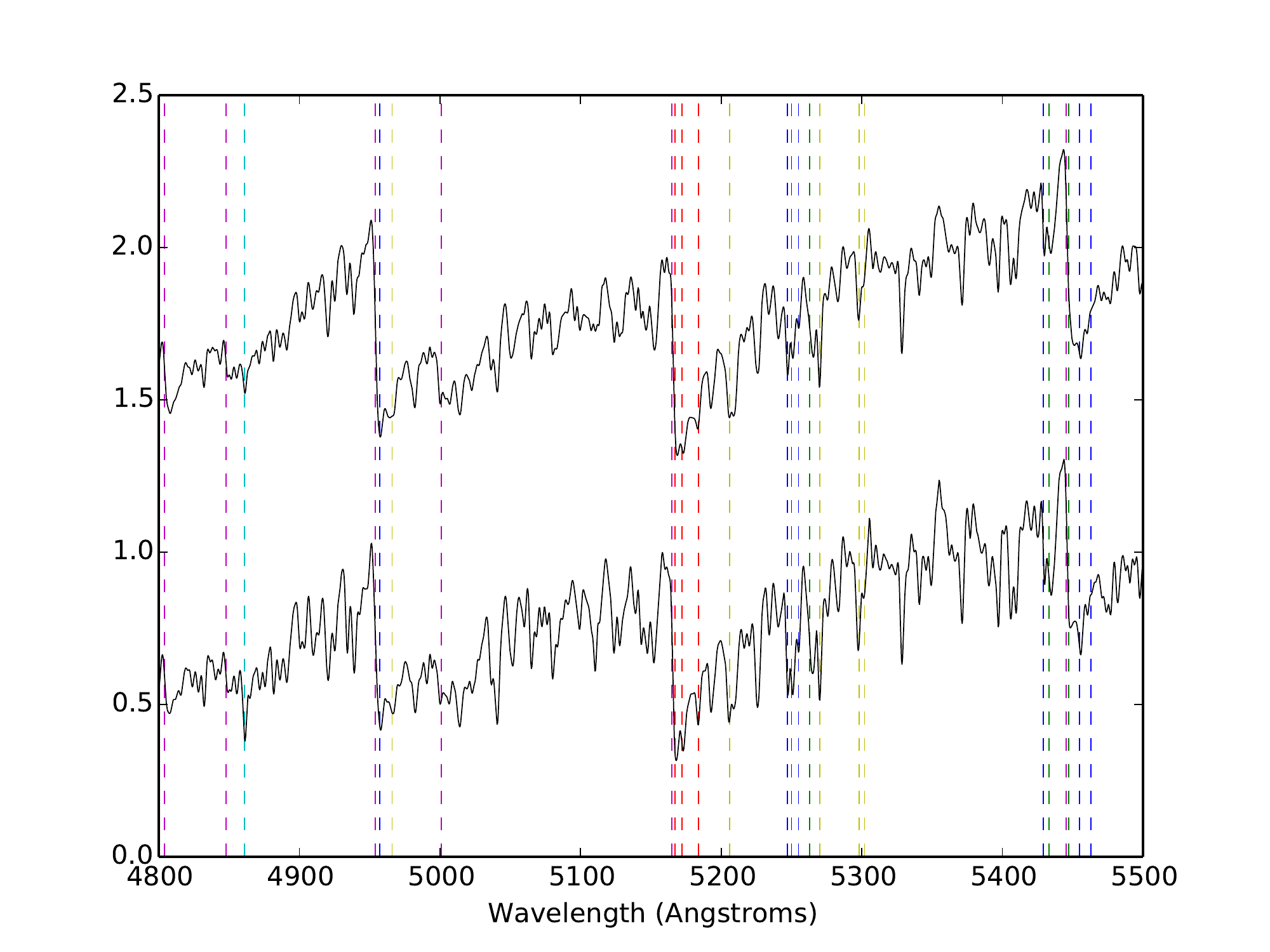}
   \caption{Example of the effect of luminosity in early- to mid-M stars. The stars shown are, from bottom to top, HD~42543 (M1\,--\,M2\:Ia\,--\,Iab) and HD~217906 (M2.5\:II\,--\,III). Dashed lines are as in Fig.~\ref{MgT_G}.}
   \label{MgT_M_2}
\end{figure*}

It is important to  note that the spectral classification described in this work is a morphological classification performed by visually comparing our spectra with those of MK~standard stars. We do not make any assumptions about the physics of the stellar interior or about the physical parameters of the atmosphere. Although there is a good correlation in general terms between spectral type and physical characteristics, this is not always true. This statement is especially important when dealing with luminous and cool stars because intermediate-mass  stars ($<8\,$M$_{\odot}$) in the AGB phase may reach luminosities similar to those of low-luminosity high-mass stars ($>10\,$M$_{\odot}$). A prime example of this is offered by $\alpha$~Her, an M5\,Ib\,--\,II MK standard. This is an important star for classification because it is the high-luminosity standard with the latest SpT. However, \citet{mor2013} have presented evidence that $\alpha$~Her is an AGB star of only $\sim3\:$M$_{\odot}$, not a high-mass star.

\subsection{Classification at $\lambda/\delta\lambda\sim1\,300$}
\label{cl_580}

Spectra observed with the 580V grating cover roughly from $3730\:$\AA{} to $5850\:$\AA{} (the exact limits depend on fibre position). However, the S/N bluewards of $\sim4500\:$\AA{} is very low for many of our stars (though not for the earlier spectral types). In consequence, most of our classification criteria lie between $4500\:$\AA{} and $5850\:$\AA{}.

We employed as main LC indicators the ratios between the different lines of the Mg\,{\sc{i}} triplet (MgT onward), whose wavelengths are 5167\:\AA{}, 5172\:\AA{}, and 5184\:\AA{} \citep{fit1951}. From G0 up to $\sim$M3, Mg\,{\sc{i}}~5167\:\AA{} is clearly deeper than the other two lines in LC~I stars (see Figs.~\ref{MgT_G}, \ref{MgT_G_2}, \ref{MgT_K}, \ref{MgT_M}, and~\ref{MgT_M_2}). When the three lines present similar intensities (although Mg\,{\sc{i}}~5167\:\AA{}  tends to be  slightly deeper than the other two), the star is classified as LC~II. Finally, if Mg\,{\sc{i}}~5184\:\AA{} is deeper than Mg\,{\sc{i}}~5167\:\AA{} the star is a giant (LC~III) or a less luminous star. These ratios change slowly with SpT: from G0 down to early -M subtypes the lines Mg\,{\sc{i}}~5172\:\AA{} and Mg\,{\sc{i}}~5184\:\AA{} gradually increase their relative intensity with respect to Mg\,{\sc{i}}~5167\:\AA{}. This SpT-dependent behaviour introduces some uncertainty for the least luminous SGs (LC~Ib) and the most luminous giants (LC~II), but its effect is not severe enough to complicate the identification of mid- and high-luminosity SGs (Iab, Ia).

\begin{figure*}
   \centering
   \includegraphics[trim=1.8cm 0.4cm 1.5cm 1.3cm,clip,width=12.5cm]{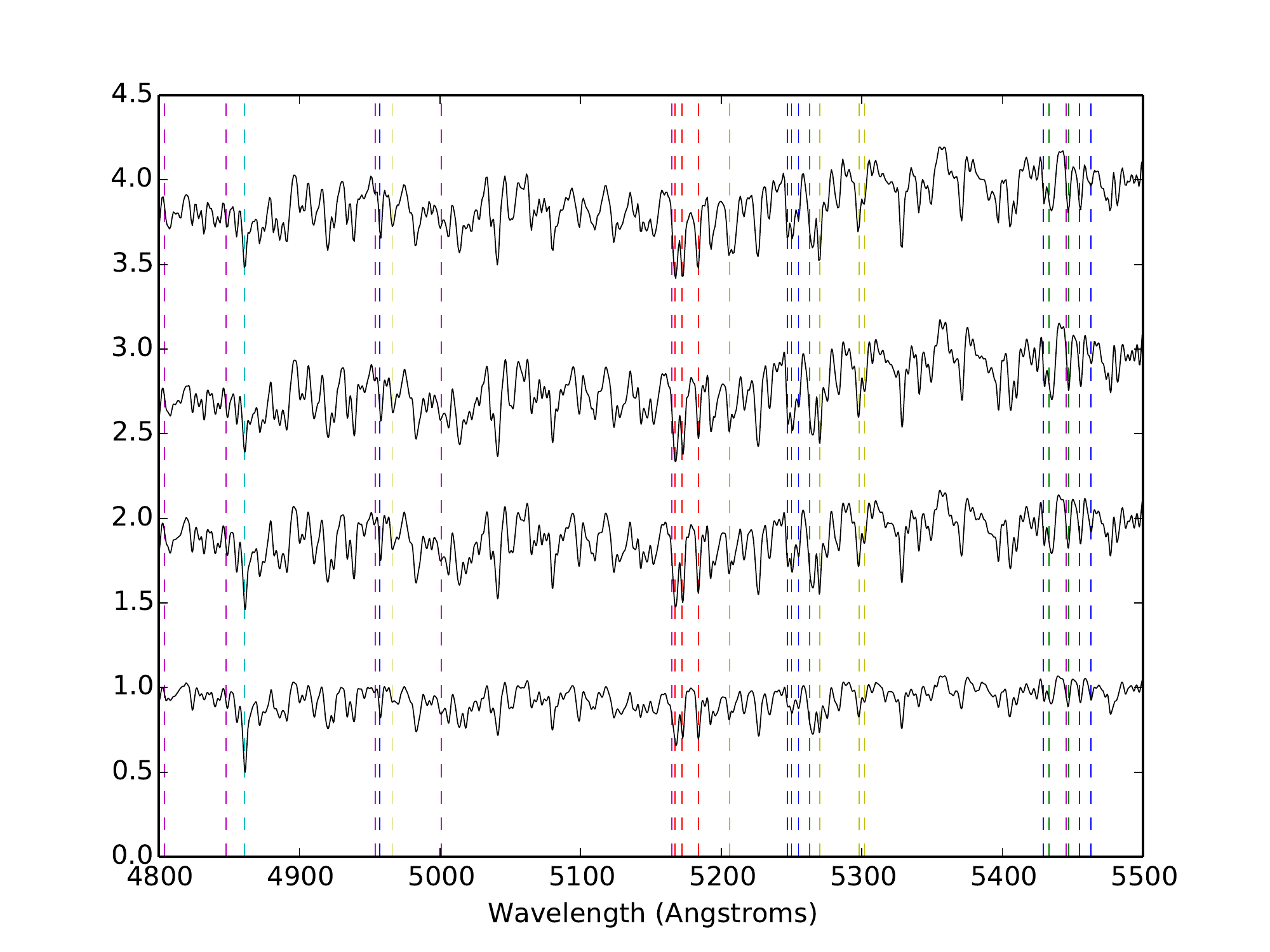}
   \caption{Example of a SpT sequence for supergiants along G and early-K subtypes. The stars shown are, from bottom to top, HD~74395 (G1\:Ib), HD~20123 (G5\:Ib\,--\,IIa), HD~48329 (G8\:Ib), and HD~63302 (K1\:Ia\,--\,Iab). Dashed lines are as in Fig.~\ref{MgT_G}.}
   \label{spt_seq_G}
\end{figure*}

\begin{figure*}
   \centering
   \includegraphics[trim=1.8cm 0.4cm 1.5cm 1.3cm,clip,width=12.5cm]{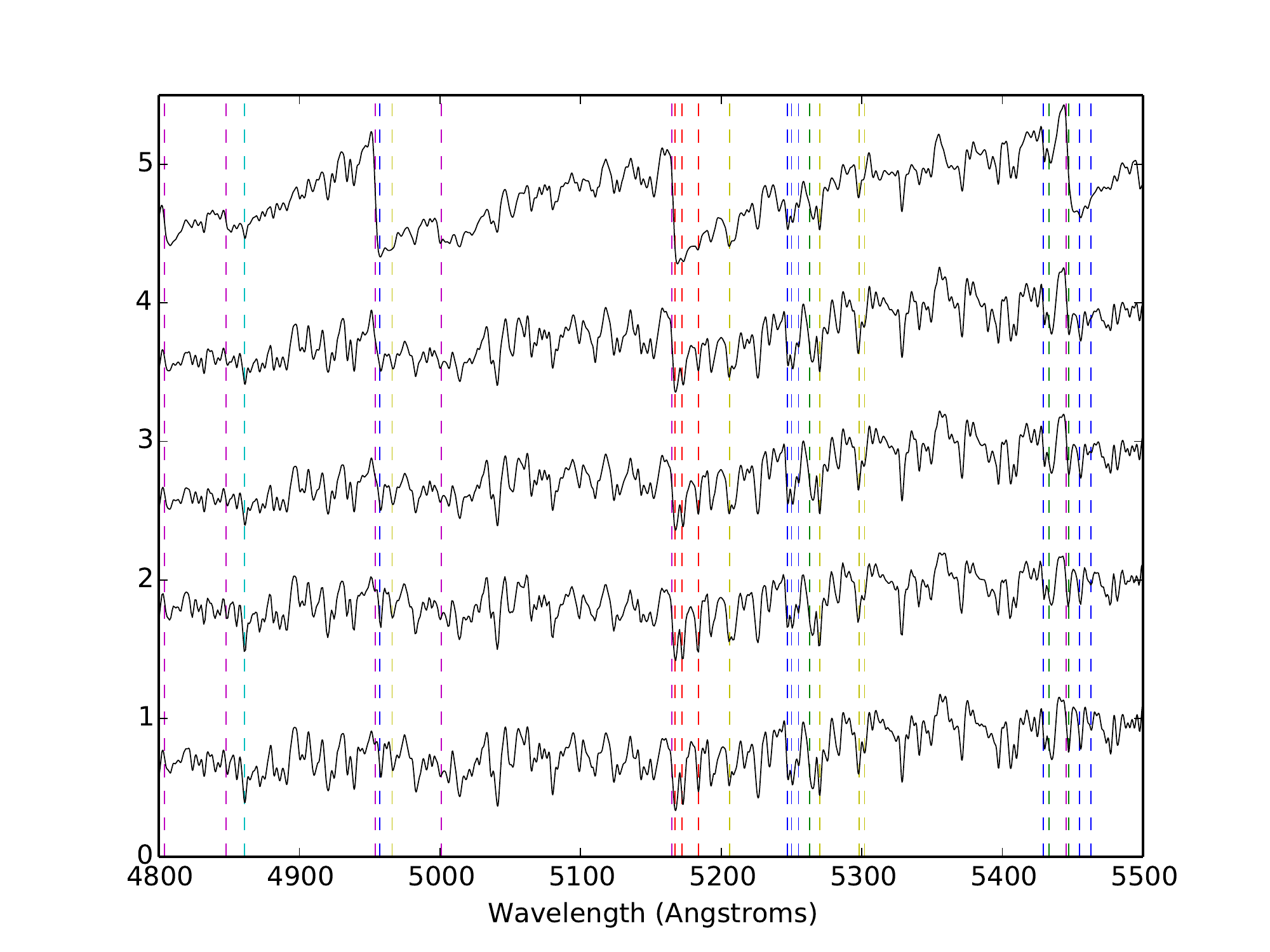}
   \caption{Example of a SpT sequence for luminous stars along K and early-M subtypes. The stars shown are, from bottom to top, HD~63302 (K1\:Ia\,--\,Iab),  HD~52005 (K3\:Ib), HD~44537 (K5\,--\,M1\:Ia\,--\,Iab), HD~206936 (M2-\:Ia), and HD~175588 (M4\:II). Dashed lines are as in Fig.~\ref{MgT_G}.}
   \label{spt_seq_K_M}
\end{figure*}

A number of other spectral features can be used to confirm the LC derived from the MgT, but they require some  knowledge of the SpT. Two features are particularly useful for a first estimation of the SpT (see Figs.~\ref{spt_seq_G} and~\ref{spt_seq_K_M}); more specific criteria  for giving a more precise classification are discussed later. Firstly, for early- and mid-G stars there is an unidentified line at $\sim$5188\:\AA{}. It is deep at early-G~subtypes, but quickly becomes weaker for later subtypes. Secondly, there is a TiO bandhead at 5167\:\AA{}, whose growth with SpT changes the shape of the MgT and the nearby continuum. This band is first noticed at K1 as a depression of the continuum between Mg\,{\sc{i}}~5167\:\AA{} and Mg\,{\sc{i}}~5172\:\AA{}. At late-K subtypes, it clearly affects the whole MgT, decreasing the depth of its lines, and also the nearby continuum. For M3 or later spectral types, the MgT is useless for LC classification because of the effect of the TiO band on it. Based on these general behaviours, we can set the following rules:

\begin{itemize}
\item[-] If the atomic line at $\sim$5188\:\AA{} is clearly noticeable, the SpT is earlier than G7.
\item[-] If the line at $\sim$5188\:\AA{} is almost not noticeable, but the continuum between Mg\,{\sc{i}}~5167\:\AA{} and Mg\,{\sc{i}}~5172\:\AA{} is not clearly affected by TiO~5167\:\AA{}, the SpT is between G7 and $\sim$K1.
\item[-] When the MgT lines and the TiO~5167\:\AA{} bandhead are all noticeable, the SpT is between $\sim$K1 and $\sim$M3.
\item[-] If the lines of the MgT are so affected by the TiO that they are no longer useful for LC classification,  the SpT is M3 or later.
\end{itemize}

Once the SpT is approximately determined, we can confirm the LC (see Figs.~\ref{MgT_G}, \ref{MgT_G_2}, \ref{MgT_K}, \ref{MgT_M}, and~\ref{MgT_M_2}) with other criteria:

\begin{itemize}
\item[-] The ratio between the group formed by three Fe\,{\sc{i}} lines (at 5247\:\AA{}, 5250\:\AA{}, and 5255\:\AA{}), and the blend at 5270\:\AA{} (formed by lines of Ca\,{\sc{i}}, Fe\,{\sc{i}}, and Ti\,{\sc{i}}) is indicative of the LC for a broad range of SpTs \citep{fit1951}. While the blend at 5270\:\AA{} hardly depends on luminosity, the Fe\,{\sc{i}} group around 5250\:\AA{} is very sensitive to it. There are two limitations to the use of this ratio. Firstly, it is useless for early- and mid-G stars because the Fe\,{\sc{i}} blend is so weak at these subtypes that its changes between different LCs are too small. Secondly, the Fe\,{\sc{i}} blend grows slowly with SpT, thus it is important to have an estimation of it (through the MgT) prior to comparing the lines.

\item[-] The ratio of Mn\,{\sc{i}}~5433\:\AA{} to Mn\,{\sc{i}}~5447\:\AA{} is also indicative of the LC. For stars earlier than K1, Mn\,{\sc{i}}~5433\:\AA{} is more intense than Mn\,{\sc{i}}~5447\:\AA{} at LC~I, similar at Ib\,--\,II, and slightly fainter for LC~III. In stars with K or early-M subtypes, the intensity of the two lines is similar even for Iab stars. This criterion becomes useless for subtypes M3 or later because of the growth of a TiO band at 5447\:\AA{}. 

\item[-] When the spectral region from 4100\:\AA{} to 4500\:\AA{} had a high enough S/N, we used the ratio of Fe\,{\sc{i}}+Y\,{\sc{ii}} at 4376\:\AA{} to Fe\,{\sc{i}} at 4383\:\AA{} \citep{kee1976}. For LC~III, Fe\,{\sc{i}}~4376\:\AA{} is more intense than the other, but at LC~I it is similar or just slightly weaker. This ratio is useful for all SpTs earlier than $\sim$M5.
\end{itemize}

Once the LC of a given star was known and its SpT roughly estimated, we determined with more precision its spectral subtype and, when the stars was a SG, its luminosity subclass (Ia, Iab, or Ib).

If the SpT was estimated earlier than G7 (see Fig.~\ref{spt_seq_G}), we used the following criteria. The Balmer lines decrease in strength along the SpT sequence for the range we handle, while at the same time metallic lines become more intense. Thus, for the identification of G and earlier subtypes we used the ratio of H$\beta$ and H$\gamma$ to other nearby metallic lines. For those stars with a high enough S/N in this region, we also compared H$\gamma$ to the G band (mainly formed by CH absorption, it spans from 4290\:\AA{} to 4314\:\AA{}): F stars have a H$\gamma$ deeper than the G band; at G0 the two features have similar depths; from that subtype down to mid-G the G band becomes dominant. When we could not use these features because of the low S/N on the blue side of our spectral range, we used four others at redder wavelengths (but with reduced discriminating power): 1) the ratio of H$\beta $, which has a similar behaviour to H$\gamma$, to nearby metallic lines; 2)  the ratio of the line at $\sim$5188\:\AA{}  to the MgT lines, because this line weakens along G~subtypes; 3)  the line Cr\,{\sc{i}}+Mn\,{\sc{i}} at $\sim$5298\:\AA{} is similarly weak to the Fe\,{\sc{i}}+Cr\,{\sc{i}} line at 5302\:\AA{} in F\,I stars, but is clearly deeper in G0\,I or later stars; and 4)  the lines Fe\,{\sc{i}}~5429\:\AA{} and Mn\,{\sc{i}}~5433\:\AA, which{} are similar to Fe\,{\sc{i}}~5424\:\AA{} in G0\,I stars but  grow towards later SpTs faster than the Fe\,{\sc{i}}~5424\:\AA{} line does. In consequence, at $\sim$K1~I the first two lines are approximately twice as deep as the other one.

For late-G and early-K subtypes, the changes in the spectra along the SpT sequence are subtle (see Fig.~\ref{spt_seq_G}). We used as criteria the progressive weakening of the $\sim$5188\:\AA{} line until it becomes unnoticeable at $\sim$K1, and the small changes in shape of the continuum between Mg\,{\sc{i}}~5167\:\AA{} and Mg\,{\sc{i}}~5172\:\AA{}: although the TiO band at 5167\:\AA{} is only clearly noticeable beyond K1, it is possible to observe the variation in this continuum from G7 to K1 as the band grows.

The subtypes from early-K up to early-M were identified attending to the changes in the shape of the metallic lines caused by the rise of TiO bands (see Fig.~\ref{spt_seq_K_M}). The effects of these bands over the spectra are noticeable from K0 onwards, and the sequence of M~subtypes is defined by the depth of their bandheads \citep{tur1985}. There are two TiO bands whose rise becomes noticeable at K1, one at 5167\:\AA{} and the other at 5447\:\AA{}. The first  changes the shape of the continuum between the MgT lines in a very apparent way (as explained above). The second  affects the lines Mn\,{\sc{i}}~5447\:\AA{} and Mn\,{\sc{i}}~5455\:\AA{}, whose depth ratio is indicative of the SpT. For G and early-K subtypes, Mn\,{\sc{i}}~5447\:\AA{} is deeper than Fe\,{\sc{i}}~5455\:\AA{}. This ratio begins to change at K3, despite the growth of the TiO~5447\:\AA{} band from $\sim$K2 onward or maybe because of it. At K4 the depths of the two lines become equal, and at K5, Fe\,{\sc{i}}~5455\:\AA{} is deeper than Mn\,{\sc{i}}~5447\:\AA{}. When the LC is III, the TiO~5447\:\AA{} band becomes noticeable at later subtypes ($\sim$K3) than in the case of LC~I. The K5 subtype can also be identified through the TiO band at 4954\:\AA{}. It affects the continuum between the  Fe\,{\sc{i}}~4957\:\AA{} line and the blend of Cr\,{\sc{i}} and Co\,{\sc{i}} at 4966\:\AA{}. The intensity of the continuum in this region decreases because of the bandhead, and by K5 its depth is half that of the atomic lines around it. The transition from K5 to M2 subtypes can be identified using the depth of the TiO bandheads at 4954\:\AA{}, 5167\:\AA{}, and 5447\:\AA{} as they grow for later subtypes.

For the spectral classification of the mid- and late-M subtypes we used a number of TiO bandheads because the appearance of each one is characteristic of a given subtype \citep{tur1985}: 

\begin{itemize}
\item[-] M2: The main criterion is the appearance of the TiO band at 4761\:\AA{}. There are three other  TiO bands useful for  identifying this subtype,  5598\:\AA{}, 5629\:\AA{}, and 5661\:\AA{}, but they are outside the spectral range observed for most of our stars;
\item[-] M3: The appearance of the TiO bands at 4804\:\AA{} and 5003\:\AA{};
\item[-] M4: The appearance of the TiO bands at 4584\:\AA{}, 4626\:\AA{}, and especially 4848\:\AA\  because it is easier to identify;
\item[-] M5: For this subtype there is only one indicative TiO band at 5569\:\AA{}, but it was not inside the  spectral range observed in all our stars.
\end{itemize}

\begin{figure*}
   \centering
   \includegraphics[trim=1.8cm 0.4cm 1.5cm 1.3cm,clip,width=12.5cm]{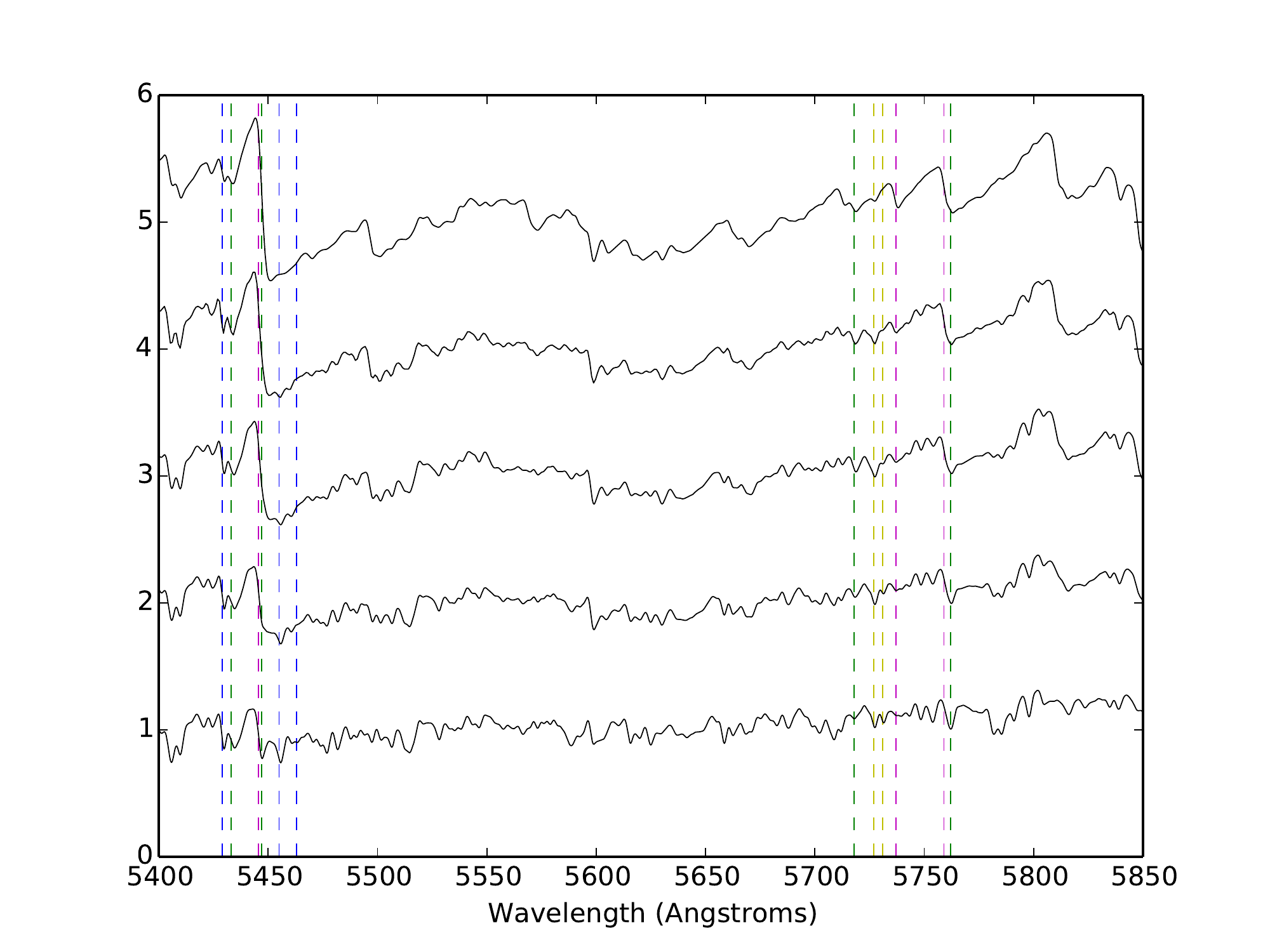}
   \caption{Example of a SpT sequence for giants along M subtypes. We use giants instead supergiants because there are no mid- or late-M standards with LC~I. The stars are shown  from bottom to top:  HD~168720 (M1\:III), HD~167006 (M3\:III), HD~175588 (M4\:II), HD~172380 (M4.5\,--\,M5\:II), and HD~196610 (M6\:III). Dashed lines are as in Fig.~\ref{MgT_G}.}
   \label{spt_seq_M}
\end{figure*}

We also used the spectral features located between 5700\:\AA{} and 5800\:\AA{} when available as not all the spectra observed include this whole range. This region contains many lines and bands useful to classify mid- and late-M subtypes (see Fig.~\ref{spt_seq_M}). The line V\,{\sc{ii}}~5731\:\AA{} is clearly present in all K~subtypes, but its strength decreases quickly along early-M subtypes: at M1\,--\,M2 the line is very weak and at M3 it is almost gone. The line Mn\,{\sc{i}}~5718\:\AA{} shows  the opposite behaviour. It is not noticeable before early-M subtypes, but it grows quickly along the M sequence. It is an intense line at M2\,--\,M3, but not as deep as the line V\,{\sc{i}}~5727\:\AA{}. Because of the appearance of an unidentified band in this region at $\sim$M4.5, the two lines (Mn\,{\sc{i}}~5718\:\AA{} and V\,{\sc{i}}~5727\:\AA{}) reach roughly the same depth. At $\sim$M5, Mn\,{\sc{i}}~5718\:\AA{} becomes deeper than the other. The M5\,--\,M5.5 subtypes can be identified by the VO band at 5737\:\AA{} as its bandhead reaches a similar depth to that of V\,{\sc{i}}~5727\:\AA{}. Finally at M6 the VO band becomes deeper than the V~line, and similar to Mn\,{\sc{i}}~5718\:\AA{}. The TiO~5759\:\AA{} band is first noticed at M0. Its growth affects the Ti\,{\sc{i}}~5762\:\AA{} line, and also the continuum at the red side of the Ti~line. At M2 this continuum is depressed down to  half the depth of the Ti~line, and at M4 the line has fully disappeared because of the band. At M6 the bandhead  of VO~5737\:\AA{} has almost the same depth as TiO~5759\:\AA{}, becoming clearly deeper than it at M7, and doubling the depth of the TiO bandhead at M8.

\subsection{Classification at $\lambda/\delta\lambda\sim3\,700$}
\label{cl_1500}

\begin{figure*}
   \centering
   \includegraphics[trim=1.8cm 0.4cm 1.5cm 1.3cm,clip,width=12.5cm]{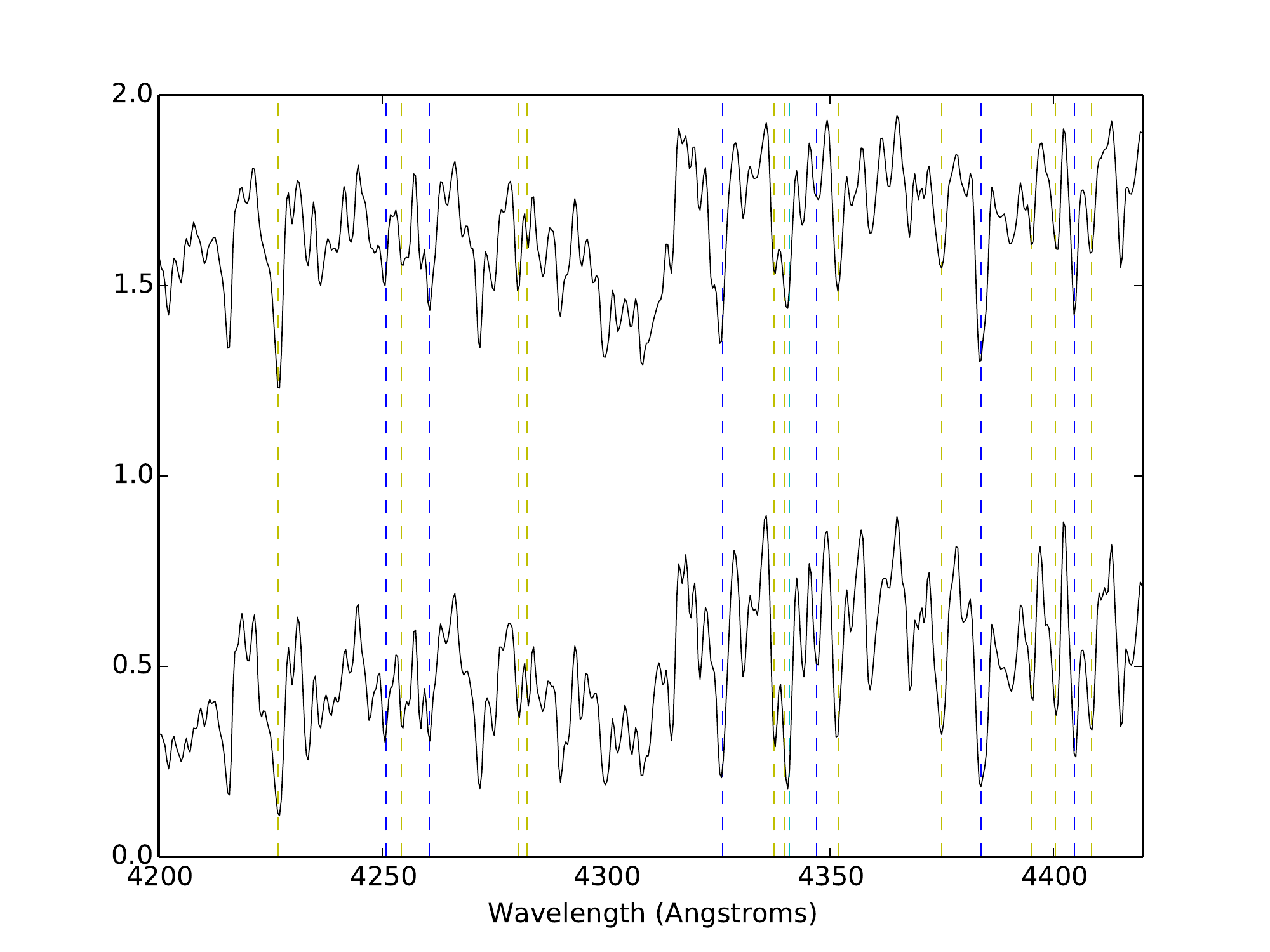}
   \caption{Example of the effect of luminosity in late-G stars. The stars are shown  from bottom to top: HD~48329 (G8\:Ib) and HD~10761 (G8\:III). Dashed lines are as in Fig.~\ref{MgT_G}.}
   \label{HR_G}
\end{figure*}

\begin{figure*}
   \centering
   \includegraphics[trim=1.8cm 0.4cm 1.5cm 1.3cm,clip,width=12.5cm]{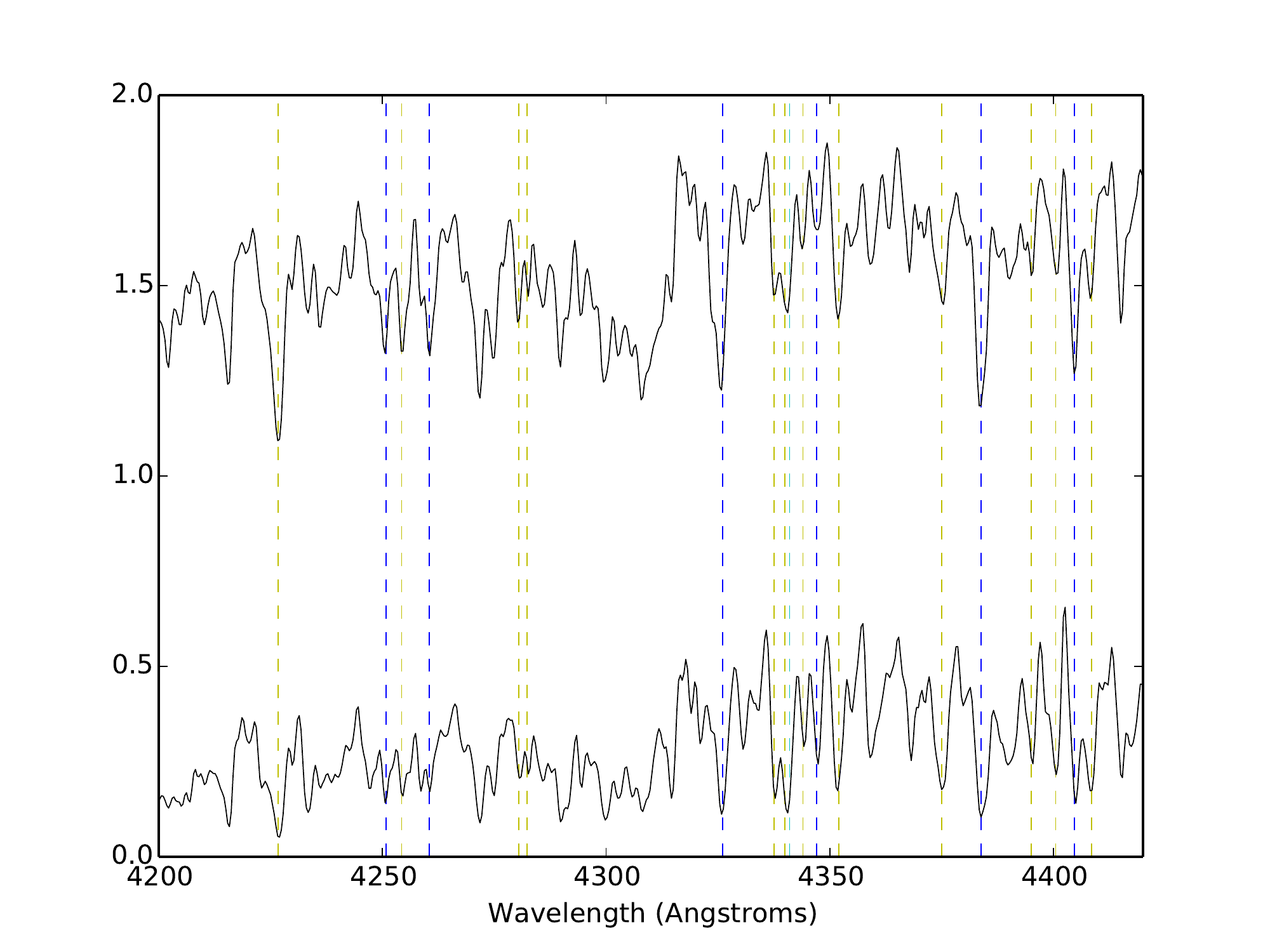}
   \caption{Example of the effect of luminosity in early-K stars. The stars are shown  from bottom to top:  HD~63302 (K1\:Ia\,--\,Iab) and HD~28292 (K1\:IIIb). Dashed lines are as in Fig.~\ref{MgT_G}.}
   \label{HR_K}
\end{figure*}

\begin{figure*}
   \centering
   \includegraphics[trim=1.8cm 0.4cm 1.5cm 1.3cm,clip,width=12.5cm]{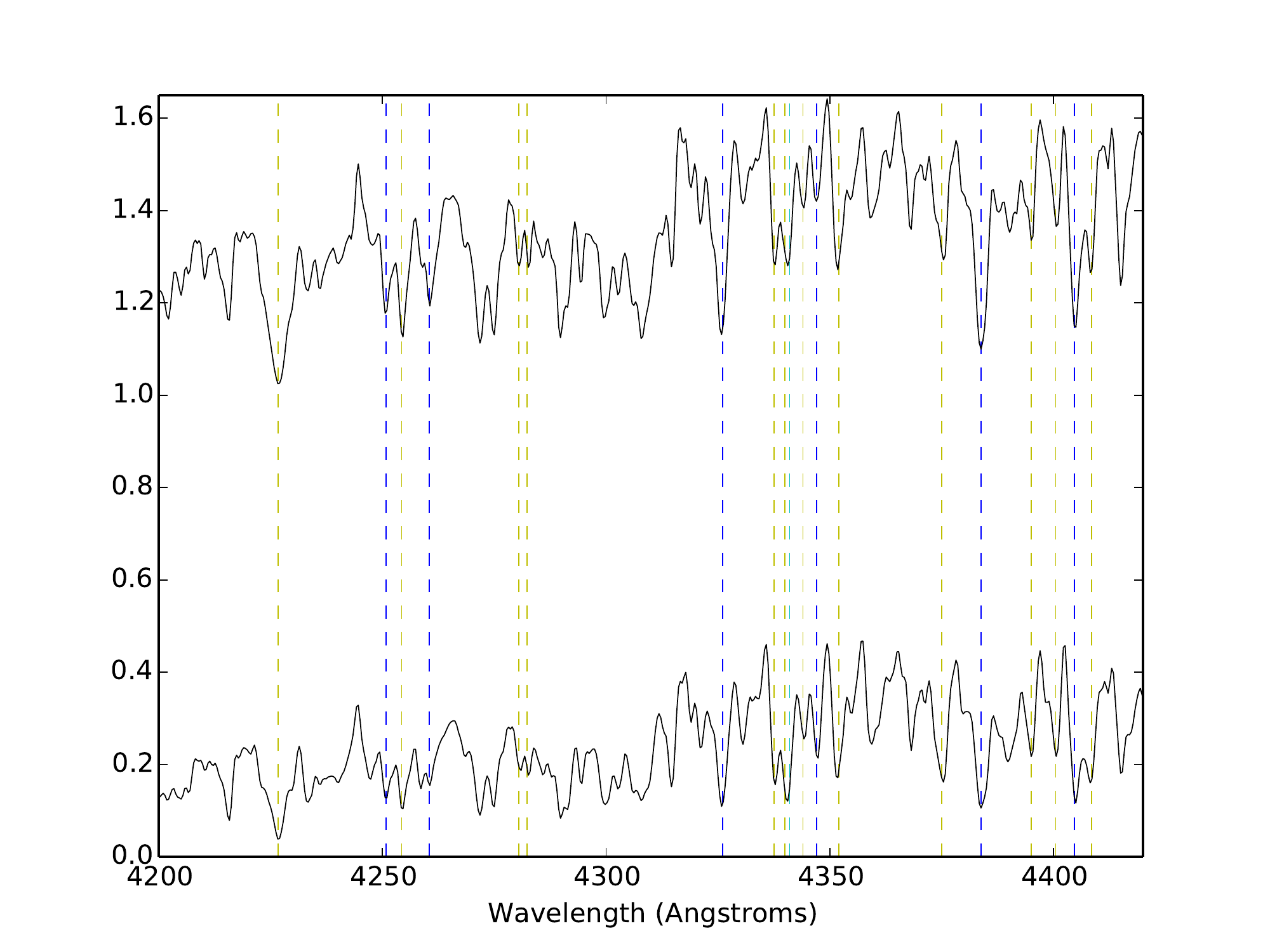}
   \caption{Example of the effect of luminosity in late-K and early-M stars. The stars are shown from bottom to top: HD~44537 (K5\,--\,M1\:Iab\,--\,Ib) and HD~113996 (K5-\:III). Dashed lines are as in Fig.~\ref{MgT_G}.}
   \label{HR_M}
\end{figure*}

For spectra observed with the 1500V grating, which provides a higher resolution but covers a shorter spectral range (see Figs.~\ref{HR_G}, \ref{HR_K}, and~\ref{HR_M}), we used different criteria, although the methodology was the same. We identified the LC using the following ratios from \cite{kee1976}: the blend of Fe\,{\sc{i}} and Sr\,{\sc{ii}} at 4216\:\AA{} to the Ca\,{\sc{i}} line at 4226\:\AA{}, the blend of Fe\,{\sc{i}} and Y\,{\sc{ii}} at 4374.5\:\AA{} to the Fe\,{\sc{i}} line at 4383\:\AA{}, and the line of Fe\,{\sc{i}} at 4404\:\AA{} to the blend of Fe\,{\sc{i}}, V\,{\sc{i}}, and Ti\,{\sc{ii}} at 4409\:\AA{}. In all cases, the ratios are $\sim$1 in LC~I and $\gg$1 for less luminous stars (LC~III\,--\,V).

The SpT can be evaluated by comparing H$\delta$ at 4102\:\AA{} and H$\gamma$ at 4341\:\AA{} to nearby metallic lines. For F or earlier subtypes H$\gamma$ is clearly dominant, while for early-G subtypes the depth of H$\gamma$ has decreased, and it is similar to that of the G band. However, this decrease stops at mid-G because of the growth of metallic lines blended with H$\gamma$.

\begin{figure*}
   \centering
   \includegraphics[trim=1.8cm 0.4cm 1.5cm 1.3cm,clip,width=12.5cm]{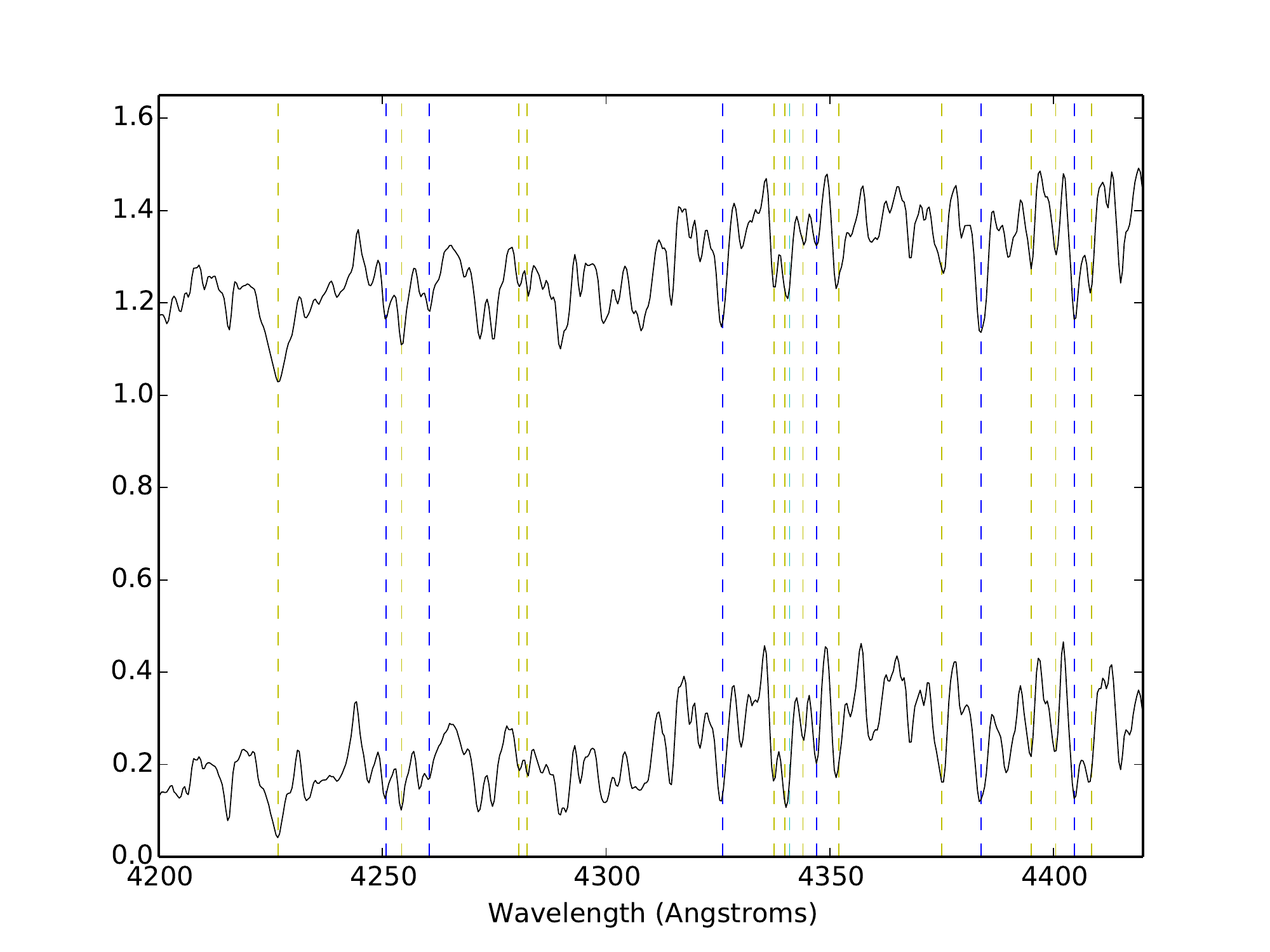}
   \caption{Example of the effect of luminosity in early- to mid-M stars. The stars are shown from bottom to top: HD~42543 (M1\,--\,M2\:Ia\,--\,Iab) and HD~113996 (M2.5\:II\,--\,III). Dashed lines are as in Fig.~\ref{MgT_G}.}
   \label{HR_M_2}
\end{figure*}

\begin{figure*}
   \centering
   \includegraphics[trim=1.8cm 0.4cm 1.5cm 1.3cm,clip,width=12.5cm]{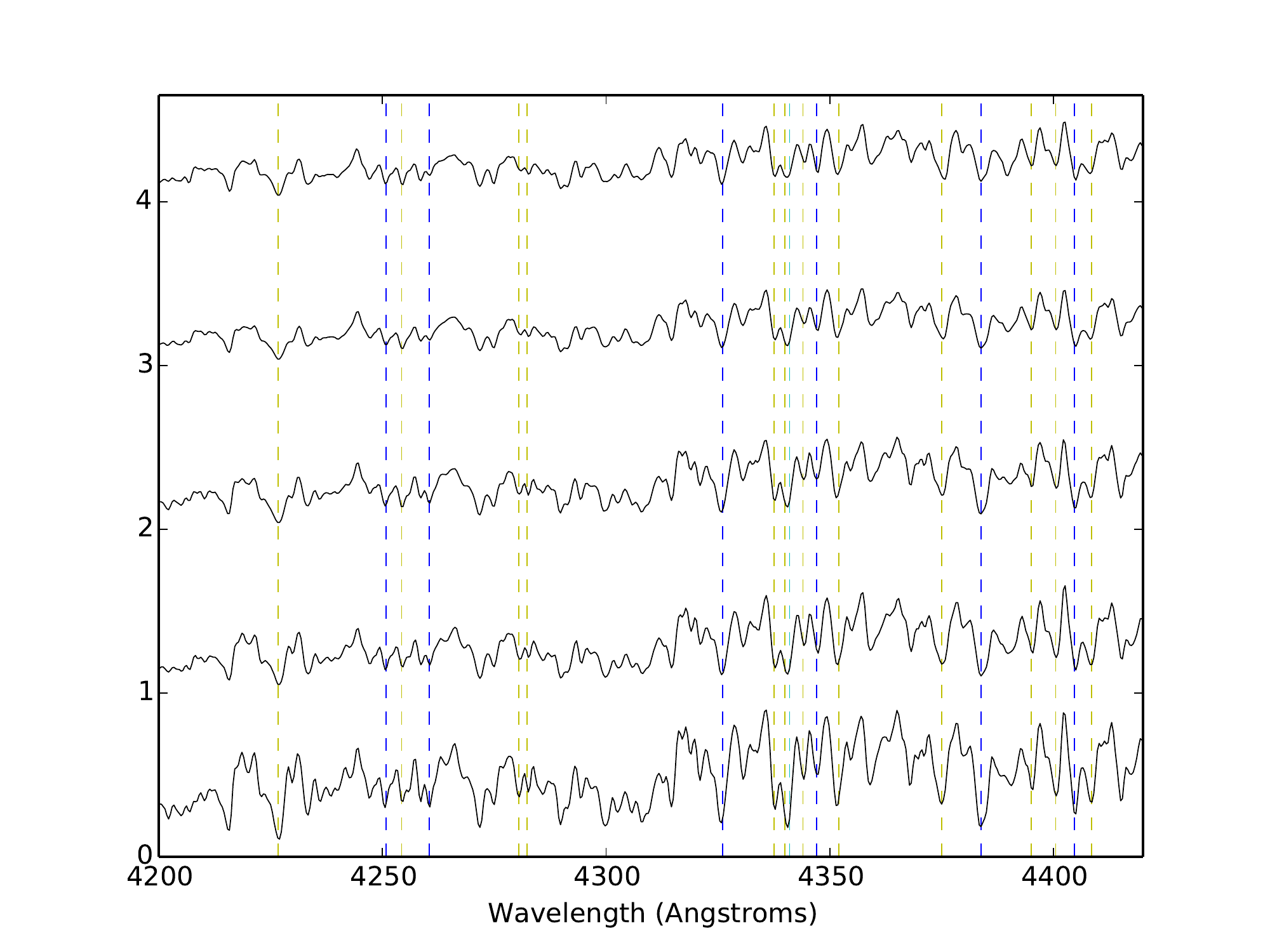}
   \caption{Example of a SpT sequence for supergiants along K and early-M subtypes. The stars
are shown  from bottom to top: HD~48329 (G8\:Ib), HD~63302 (K1\:Ia\,--\,Iab),  HD~17506 (K3-\:Ib\,--\,IIa), HD~44537 (K5\,--\,M1\:Ia\,--\,Iab), and HD~206936 (M2-\:Ia). Dashed lines are as in Fig.~\ref{MgT_G}.}
   \label{spt_seq2_K_M}
\end{figure*}

From G down to early-M subtypes (see Fig.~\ref{spt_seq2_K_M}), the blend of Fe\,{\sc{i}}, Cr\,{\sc{i}}, and Ti\,{\sc{ii}} at 4344\:\AA{} and the  Fe\,{\sc{i}} line at 4347\:\AA{} grow faster as we move to later subtypes than the blend of Mg\,{\sc{i}}, Cr\,{\sc{i}}, and Fe\,{\sc{i}} at 4351\:\AA{} does. Therefore, for early- and mid-G subtypes Fe\,{\sc{i}}~4347\:\AA{} is weaker than the blend at 4344\:\AA{}, and only for the most luminous stars (LC~Ia) do both features have similar depths. Since Fe\,{\sc{i}}~4347\:\AA{} grows deeper for later subtypes, at late-G subtypes it is about as deep as the blend, being slightly weaker in low-luminosity SGs (LC~Ib), and slightly deeper in high-luminosity SGs (LC~Ia). Even though these three lines vary with LC, we note that they can still be used to determine SpT because we can constrain the LC with other indicators before considering them. At early-K subtypes, Fe\,{\sc{i}}~4347\:\AA{} is only slightly weaker than the other two lines, while the blend at 4344\:\AA{} has a depth  between those of the two other  lines. At mid-K subtypes (K2\,--\,K4), Fe\,{\sc{i}}~4347\:\AA{} is about as deep as the blend at 4344\:\AA{} in any LC~I star. At K5 and early-M subtypes, Fe\,{\sc{i}}~4347\:\AA{} reaches the same depth as Fe\,{\sc{i}}~4351\:\AA{}, while the blend at 4344\:\AA{} is only slightly less deep than Fe\,{\sc{i}}~4351\:\AA{}.

\begin{figure*}
   \centering
   \includegraphics[trim=1.8cm 0.4cm 1.5cm 1.3cm,clip,width=12.5cm]{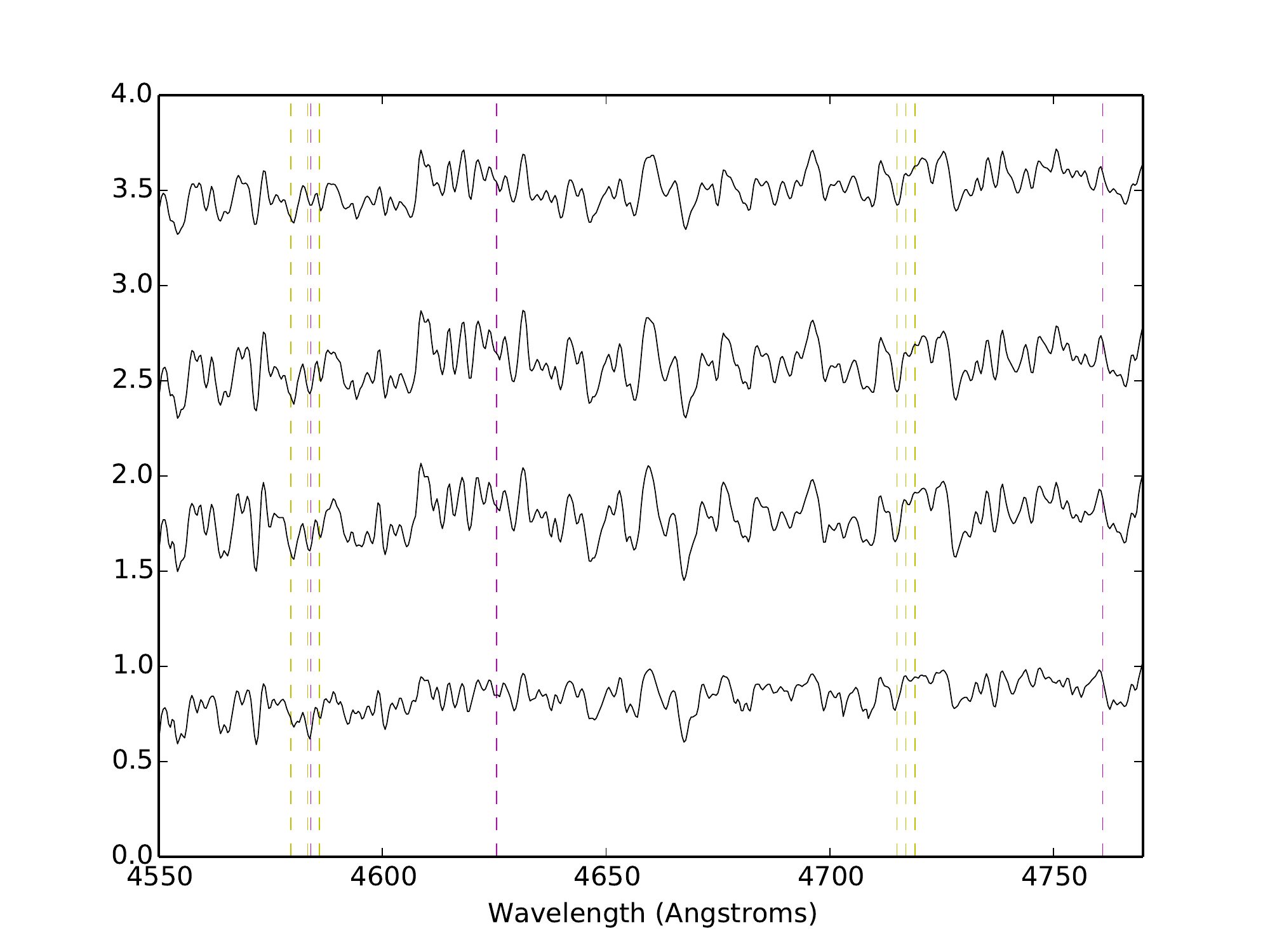}
   \caption{Example of a SpT sequence for luminous stars along G and K subtypes. The stars
are shown  from bottom to top: HD~20123 (G5\:Ib\,--\,IIa), HD~48329 (G8\:Ib), HD~63302 (K1\:Ia\,--\,Iab),  HD~17506 (K3-\:Ib\,--\,IIa), and HD~44537 (K5\,--\,M1\:Ia\,--\,Iab). Dashed lines are as in Fig.~\ref{MgT_G}.}
   \label{spt_seq3_K_M}
\end{figure*}

\begin{figure*}
   \centering
   \includegraphics[trim=1.8cm 0.4cm 1.5cm 1.3cm,clip,width=12.5cm]{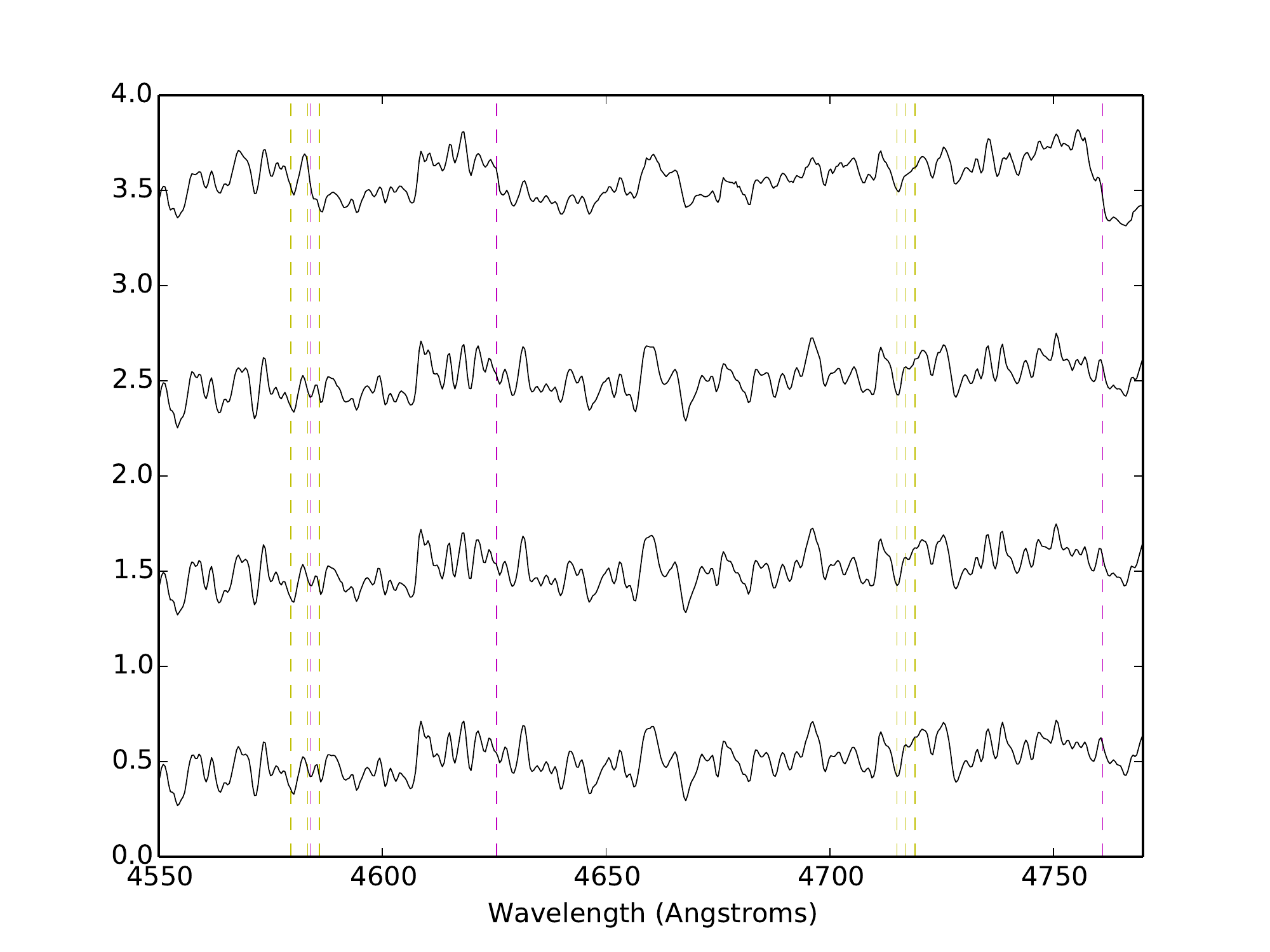}
   \caption{Example of a SpT sequence for luminous stars along late-K and M subtypes. The stars
are shown  from bottom to top: HD~44537 (K5\,--\,M1\:Ia\,--\,Iab), HD~42475 (M0\,--\,M1.5\:Iab), HD~206936 (M2-\:Ia), and HD~175588 (M4\:II). Dashed lines are as in Fig.~\ref{MgT_G}.}
   \label{spt_seq3_M}
\end{figure*}

Once we established approximately the LC and SpT, we then  used other criteria to confirm them, obtaining the definitive classification (see Figs.~\ref{spt_seq3_K_M} and~\ref{spt_seq3_M}). The behaviours of the blends of Fe\,{\sc{i}} and Co\,{\sc{i}} at 4579\:\AA{}; the line of Fe\,{\sc{i}} at 4583\:\AA{}; and Fe\,{\sc{i}}, Cr\,{\sc{i}} and Ca\,{\sc{i}}, at 4586\:\AA{} can be used to determine the subtype in the late-G and K spectral sequences. We note that the Fe\,{\sc{i}}~4583\:\AA{} line and the blend at 4586\:\AA{} are sensitive to luminosity, and thus it is necessary to know the LC before using them.  In early-G stars Fe\,{\sc{i}}~4583\:\AA{} is clearly stronger than the blends, but at G5\,--\,G6 it becomes similar to them when LC is Ia or Iab, and only slightly weaker if the LC is Ib. At K0\,--\,K1 only in the most luminous SGs (LC~Ia) is Fe\,{\sc{i}}~4583\:\AA{} still similar or slightly deeper than the blend at 4579\:\AA{}. At K2 the line is less deep than the blend. At K3, the blend at 4586\:\AA{} and Fe\,{\sc{i}}~4583\:\AA{} are similarly deep, and at K4 the blend is clearly stronger than the Fe~line. At early-M subtypes a TiO band appears at 4584\:\AA{}, and it begins to affect the Fe\,{\sc{i}}~4583\:\AA{} line and the blend at 4586\:\AA{}. Because of this, at $\sim$M2 the blend at 4586\:\AA{} seems deeper than the blend at 4579\:\AA{}. At $\sim$M3 these lines become useless due to the effect of the TiO band on them. 

We also used the ratios of Fe\,{\sc{i}}~4251\:\AA{} to  Fe\,{\sc{i}}~4254\:\AA,{} and of Fe\,{\sc{i}}~4280\:\AA{} to the blend of Fe\,{\sc{i}} and Ti\,{\sc{i}} at 4282\:\AA{}. In subtypes earlier than K2, Fe\,{\sc{i}}~4251\:\AA{} is deeper than Fe\,{\sc{i}}~4254\:\AA{}, and Fe\,{\sc{i}}~4280\:\AA{} is deeper than the blend at 4282\:\AA{}. However, at K3 both Fe\,{\sc{i}}~4254\:\AA{} and the blend at 4282\:\AA{} become slightly deeper than Fe\,{\sc{i}}~4251\:\AA{} and Fe\,{\sc{i}}~4280\:\AA,{} respectively. In the spectral range covered by the 1500V grating, the effects of the TiO bands begin at early-M subtypes. In fact, the TiO bands are not clearly observable until $\sim$M3. To classify the early-M stars, we used the TiO band effects on the spectral ranges from 4580 to 4590\:\AA{} and from 4710 to 4720\:\AA{}. At M3\,--\,M4,  TiO~4584\:\AA{}, TiO~4626\:\AA{}, and TiO~4761\:\AA{} are clearly noticeable, but the last one is not inside the spectral range observed in all our spectra.

\section{Conclusions}
In this work we present and describe an extensive atlas of luminous late-type stars. It is composed of almost $1\,500$ spectra. For each target we provide two spectra, one taken in the optical range at moderate resolution and a higher resolution one around the infrared CaT. The atlas is mainly composed of CSGs (over $1\,000$ spectra) as our survey  focused on this kind of stars. However, it also includes a large number of typical interlopers when looking for CSGs: red giants and luminous giants, carbon stars, and a few less luminous stars. For each target, we provide its fundamental data: RV, SpT, and LC, as well as its \textit{Gaia} DR2 astrometric parameters. 

In conclusion, this atlas is the largest public collection of spectra of CSGs with spectral classification. We believe that it is a powerful tool for any group interested in the study of luminous stars from the MCs (CSGs, AGB stars, and carbon stars), or in CSGs in general. This catalogue can be used to complement  other spectroscopic surveys; for example, RAVE \citep{RAVE2017} contains very few spectra of stars in the foreground to the MCs. It can also be a useful complement to \textit{Gaia} DR2: we provide RVs for many stars that lack this datum in DR2; in other cases, our RVs can be used for multi-epoch studies in combination with \textit{Gaia} data or to  spectroscopically confirm the nature of candidates to massive stars running away from the MCs. Moreover, among our observed stars there is a large number of spectra of foreground stars, which can be useful for the study of late-type dwarf stars from the Galactic disk or halo.

In addition, we describe in detail the process of spectral classification in the optical ranges that we used for our sample. This process is based on a compilation of classical criteria. These criteria have been unified for greater coherence, and their reliability when applied to modern CCD spectra has been tested. Thus, this classification method provides a up-to-date and well-tested tool for the classification of luminous late-type stars.

\section*{Acknowledgements}

The observations have been partially supported by the OPTICON project (observing proposals 2010B/01, 2011A/014, and 2012A/015), which is funded by the European Commission under the Seventh Framework Programme (FP7).
Part of the observations have been taken under service mode (proposal AO171) and the authors gratefully acknowledge the help of the AAO support astronomers. This research is partially supported by the Spanish Government Ministerio de Econom{\'i}a y Competitividad under grant AYA2015-68012-C2-2-P (MINECO/FEDER).
The work reported on in this publication has been partially supported by the European Science Foundation (ESF), in the framework of the GREAT Research Networking Programme.
This research has made use of the Simbad database, operated at CDS, Strasbourg (France). This work has made use of data from the European Space Agency (ESA) mission
{\it Gaia} (\url{https://www.cosmos.esa.int/gaia}), processed by the {\it Gaia}
Data Processing and Analysis Consortium (DPAC,
\url{https://www.cosmos.esa.int/web/gaia/dpac/consortium}). Funding for the DPAC
has been provided by national institutions, in particular the institutions
participating in the {\it Gaia} Multilateral Agreement.

\bibliographystyle{aa}
\bibliography{general}

\appendix
\onecolumn

\section{List of standard stars used}

\begin{longtab}
\begin{longtable}{c c c}
\caption{Standard stars used for the visual spectral and luminosity classifications performed over the samples from the MCs. The last column indicates the catalogue from which the spectra were taken: the Indo--US spectral library \citep{val2004} or the MILES star catalogue \citep{sbl2006}.\label{standard}}\\
\hline\hline
\noalign{\smallskip}
HD~number&MK~type&Catalogue\\
\noalign{\smallskip}
\hline
\noalign{\smallskip}
\endfirsthead
\caption{continued.}\\
\noalign{\smallskip}
HD~number&MK~type&Catalogue\\
\noalign{\smallskip}
\hline
\noalign{\smallskip}
\endhead
020902&F5\:Ib&MILES\\
016901&G0\:Ib&MILES\\
074395&G1\:Ib&MILES\\
084441&G1\:II&MILES\\
223047&G3\:Ib\,--\,II&Indo\,--\,US\\
020123&G5\:Ib\,--\,IIa&Indo\,--\,US\\
077912&G7\:IIa&Indo\,--\,US\\
099648&G7.5\:IIIa&MILES\\
048329&G8\:Ib&Indo\,--\,US\\
202109&G8\:II&MILES\\
005516&G8$-$\:III&Indo\,--\,US\\
221115&G8\:IIIa&Indo\,--\,US\\
216131&G8+\:III&Indo\,--\,US\\
221861&G9\:Ib&Indo\,--\,US\\
180711&G9\:III&Indo\,--\,US\\
016458&K0\:Ba3:&Indo\,--\,US\\
197912&K0\:IIIa&Indo\,--\,US\\
008512&K0\:IIIb&Indo\,--\,US\\
186648&K0+\:III&Indo\,--\,US\\
164349&K0.5\:IIb&MILES\\
063302&K1\:Ia\,--\,Iab&Indo\,--\,US\\
043232&K1\:III:&Indo\,--\,US\\
028292&K1\:IIIb&Indo\,--\,US\\
206778&K2\:Ib\,--\,II&MILES\\
031767&K2\:II&MILES\\
031421&K2$-$\:IIIb&Indo\,--\,US\\
143107&K2\:IIIab&Indo\,--\,US\\
169414&K2\:IIIab&Indo\,--\,US\\
081146&K2\:IIIb&Indo\,--\,US\\
012533&K2+\:IIb&Indo\,--\,US\\
004817&K2.5\:Ib\,--\,II CN-1&Indo\,--\,US\\
083618&K2.5\:III&Indo\,--\,US\\
052005&K3\:Ib&MILES\\
156283&K3\:II&MILES\\
081797&K3\:IIIa&Indo\,--\,US\\
158899&K3.5\:III&Indo\,--\,US\\
131873&K4$-$\:III&Indo\,--\,US\\
099167&K4\:III&Indo\,--\,US\\
200905&K4.5\:Ib\,--\,II&MILES\\
149161&K4.5\:III&Indo\,--\,US\\
113996&K5$-$\:III&Indo\,--\,US\\
044537&K5\,--\,M1\:Iab\,--\,Ib&Indo\,--\,US\\
029139&K5+\:III&Indo\,--\,US\\
080493&K6\:III&Indo\,--\,US\\
020797&M0\:II&Indo\,--\,US\\
189319&M0$-$\:III&Indo\,--\,US\\
132933&M0.5\:IIb&MILES\\
146051&M0.5\:III&Indo\,--\,US\\
049331&M1\:Ib\,--\,II&MILES\\
042475&M0\,--\,M1.5\:Iab&Indo\,--\,US\\
168720&M1\:III&Indo\,--\,US\\
039801&M1\,--\,M2\:Ia\,--\,ab&Indo\,--\,US\\
042543&M1\,--\,M2\:Ia\,--\,Iab&Indo\,--\,US\\
206936&M2$-$\:Ia&Indo\,--\,US\\
036389&M2\:Iab\,--\,Ib&Indo\,--\,US\\
217906&M2.5\:II\,--\,III&Indo\,--\,US\\
040239&M3\:IIb&Indo\,--\,US\\
112142&M3$-$\:III&Indo\,--\,US\\
120933&M3$-$\:IIIa&Indo\,--\,US\\
167006&M3\:III&Indo\,--\,US\\
044478&M3\:IIIab&Indo\,--\,US\\
112300&M3+\:III&Indo\,--\,US\\
121130&M3.5\:III&MILES\\
175588&M4\:II&Indo\,--\,US\\
123657&M4.5\:III&Indo\,--\,US\\
172380&M4.5\,--\,M5\:II&MILES\\
148783&M6$-$\:III&Indo\,--\,US\\
196610&M6\:III&Indo\,--\,US\\
114961&M7\:III:&Indo\,--\,US\\
126327&M7.5\,--\,M8\:III&Indo\,--\,US\\
\noalign{\smallskip}
\hline
\end{longtable}
\end{longtab}

\pagebreak
\section{List of lines measured}

\begin{table*}
\caption{Atomic lines and molecular bands used in the classification. Those lines which are blends of multiple elements are marked as MSB (multi-species blend).}
\label{lines}
\centering
\begin{tabular}{c c c | c c c}
\hline\hline
\noalign{\smallskip}
\multicolumn{2}{c}{Atomic Line}&&\multicolumn{2}{c}{Atomic Line}&\\
$\lambda$\:(\AA{})&Chem. species&Colour used in figures&$\lambda$\:(\AA{})&Chem. species&Colour used in figures\\
\noalign{\smallskip}
\hline
\noalign{\smallskip}
4102&Balmer-$\delta$&Cyan&4848&TiO band&Magenta\\
4226.7&\ion{Ca}{i}&Yellow&4861&Balmer-$\beta$&Cyan\\
4250.8&\ion{Fe}{i}&Blue&4954&TiO band&Magenta\\
4254.3&\ion{Cr}{i}&Yellow&4957&\ion{Fe}{i}&Blue\\
4260.5&\ion{Fe}{i}&Blue&4966&MSB&Yellow\\
4280.5&MSB&Yellow&5001&TiO band&Magenta\\
4282.3&MSB&Yellow&5165&TiO band&Magenta\\
4326&\ion{Fe}{i}&Blue&5167&\ion{Mg}{i}&Red\\
4340&\ion{Cr}{i}&Yellow&5172&\ion{Mg}{i}&Red\\
4337.5&MSB&Yellow&5184&\ion{Mg}{i}&Red\\
4341&Balmer-$\gamma$&Cyan&5206&MSB&Yellow\\
4344&MSB&Yellow&5247&\ion{Fe}{i}&Blue\\
4347&\ion{Fe}{i}&Blue&5250&\ion{Fe}{i}&Blue\\
4352&MSB&Yellow&5255&\ion{Fe}{i}&Blue\\
4375&MSB&Yellow&5263&\ion{Ti}{i}&Green\\
4383.8&\ion{Fe}{i}&Blue&5270&MSB&Yellow\\
4395&MSB&Yellow&5298&MSB&Yellow\\
4400.5&MSB&Yellow&5302&MSB&Yellow\\
4404.7&\ion{Fe}{i}&Blue&5429&\ion{Fe}{i}&Blue\\
4408.5&MSB&Yellow&5433&\ion{Mn}{i}&Green\\
4462&TiO band&Magenta&5447&\ion{Mn}{i}&Green\\
4579.5&MSB&Yellow&5445.5&TiO band&Magenta\\
4583.3&MSB&Yellow&5455&\ion{Fe}{i}&Blue\\
4584&TiO band&Magenta&5463&\ion{Fe}{i}&Blue\\
4585.9&MSB&Yellow&5718&\ion{Mn}{i}&Green\\
4625.5&TiO band&Magenta&5727&\ion{V}{i}&Yellow\\
4715&MSB&Yellow&5731&\ion{V}{ii}&Yellow\\
4717&MSB&Yellow&5737&VO band&Magenta\\
4719&MSB&Yellow&5759&TiO band&Magenta\\
4761&TiO band&Magenta&5762&\ion{Ti}{i}&Green\\
4804&TiO band&Magenta\\
\noalign{\smallskip}
\hline
\end{tabular}
\end{table*}

\pagebreak
\section{Sample of observed spectra}
\label{spec_sample}

In this Appendix, we show a sample of our observed supergiants in the main ranges used for classification in this paper. We only show the supergiants of intermediate luminosity because we do not have complete SpT sequences for other luminosities. As any spectral classification  must be performed using standard stars, the spectra shown are merely an illustrative example.

\begin{figure*}[hb!]
   \centering
   \includegraphics[trim=1.8cm 0.4cm 1.5cm 1.3cm,clip,width=12.5cm]{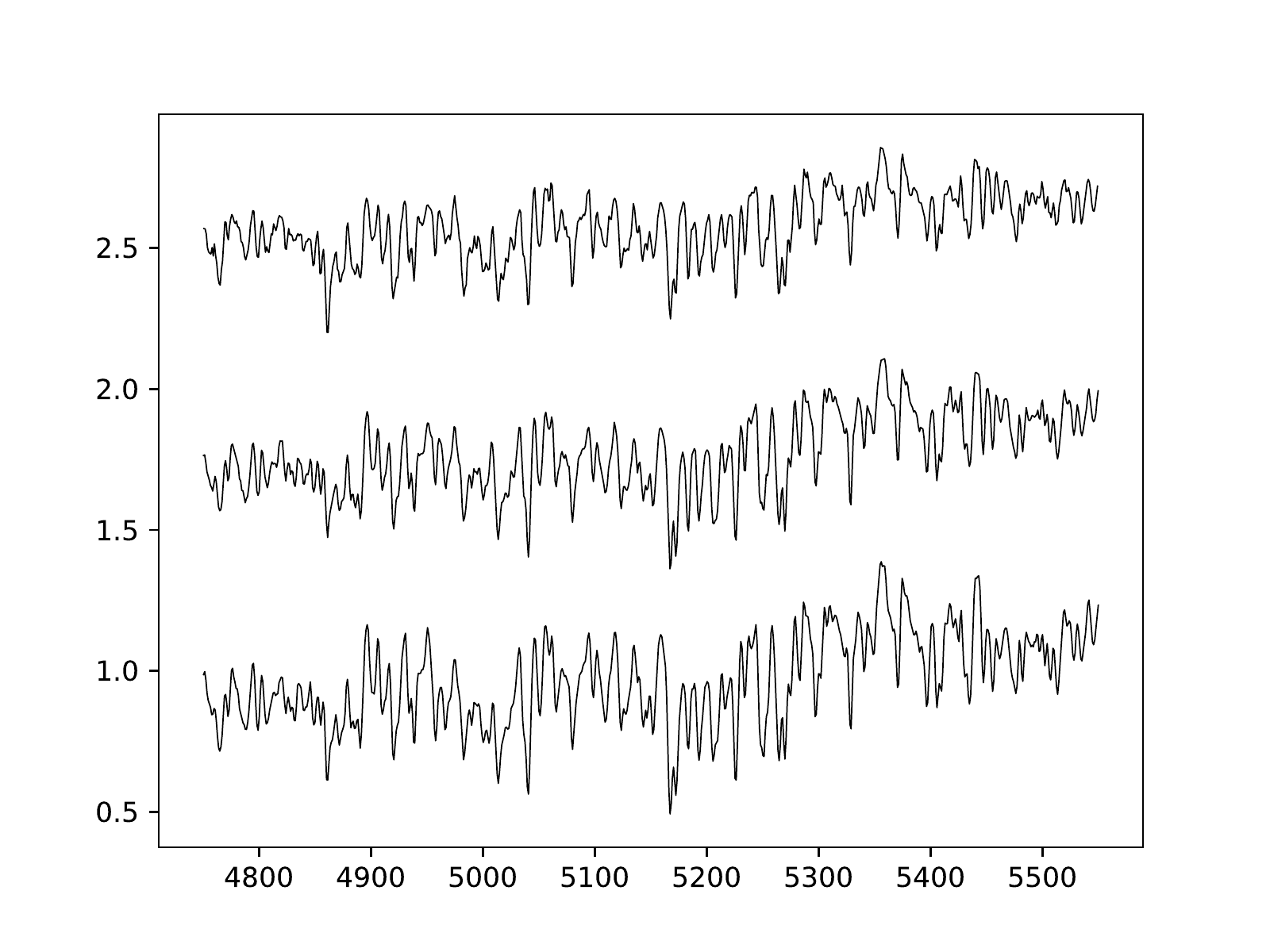}
   \includegraphics[trim=1.8cm 0.4cm 1.5cm 1.3cm,clip,width=12.5cm]{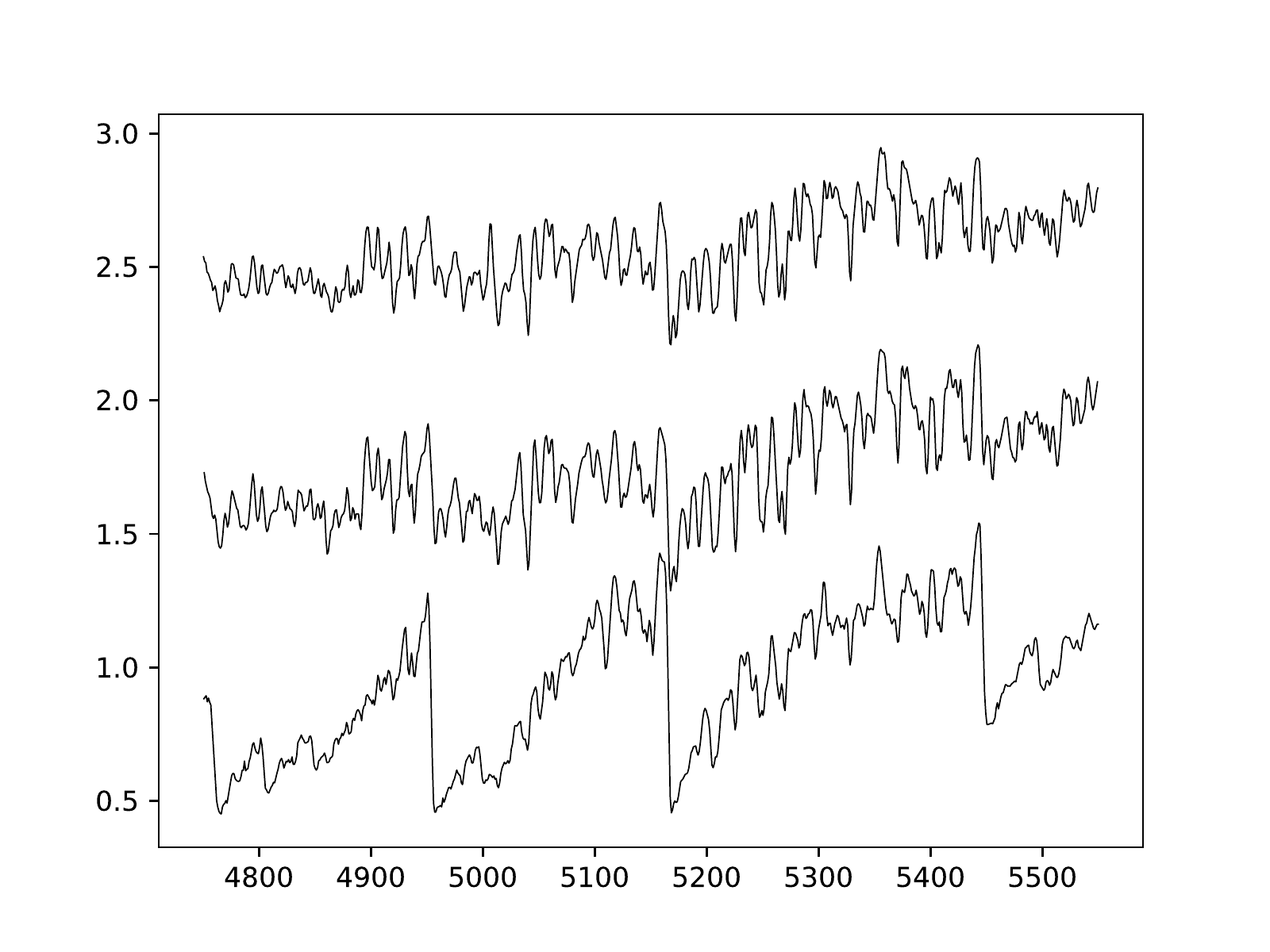}
   \caption{Spectral type sequence for supergiants from the LMC, at $\lambda/\delta\lambda\sim1\,300$. The stars are shown  from top to bottom: (upper panel) LMC035 (G5\:Iab) [2013 field-1-lmc 157], LMC164 (K2\:Iab\,--\,Ib) [2013 field-1-lmc 313], and 14576 (K5\:Iab) [2013 field-1-lmc 269]; (lower panel) 163007 (M0\:Iab) [2013 field-1-lmc 345], 130426 (M2\:Iab\,--\,Ib) [2013 field-1-lmc 206], and 135720 (M4.5\:Iab) [2013 field-1-lmc 229]. The wavelength is given in angstrom.}
   \label{LMC_LR}
\end{figure*}

\begin{figure*}
   \centering
   \includegraphics[trim=1.8cm 0.4cm 1.5cm 1.3cm,clip,width=12.5cm]{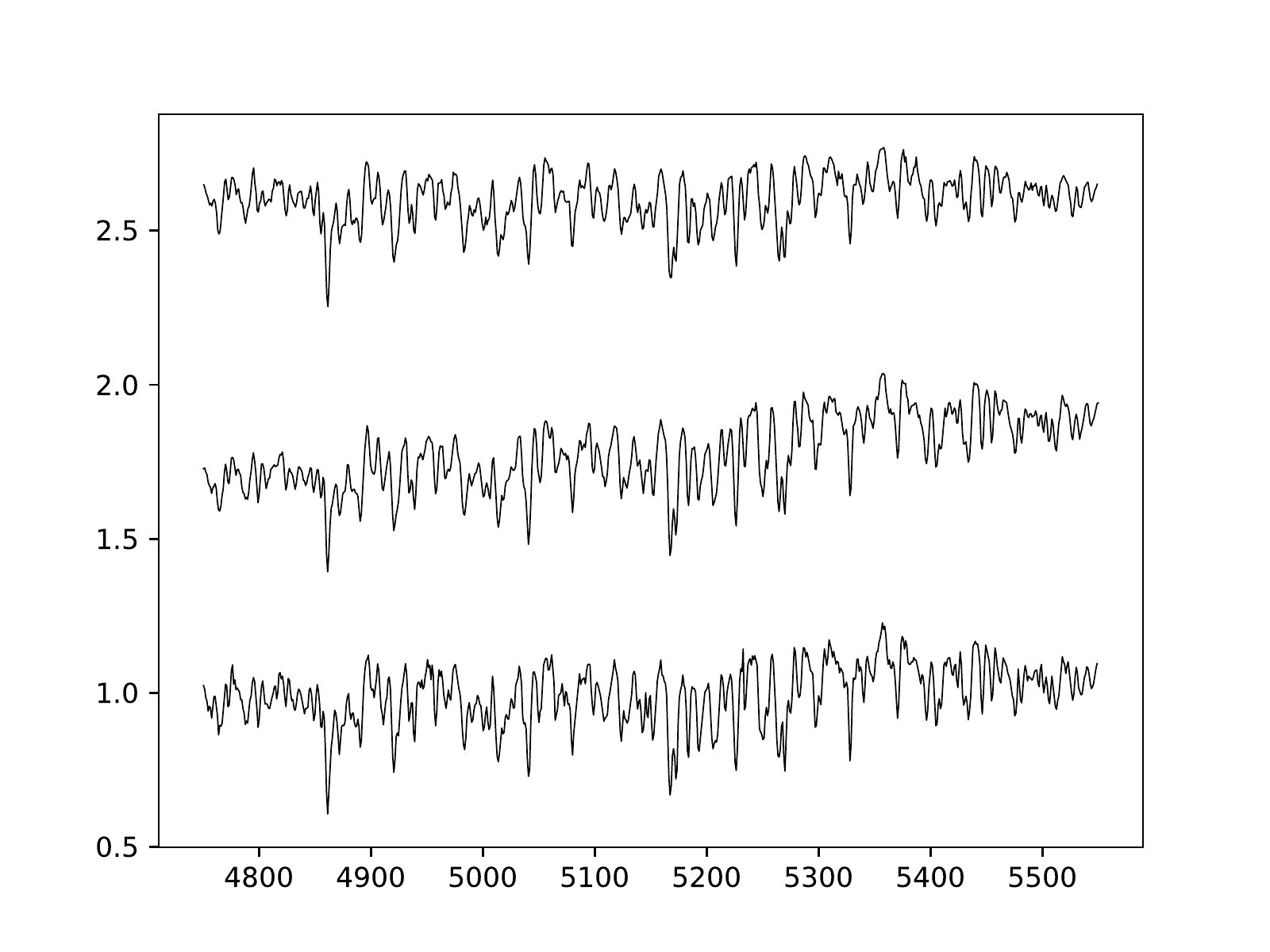}
   \includegraphics[trim=1.8cm 0.4cm 1.5cm 1.3cm,clip,width=12.5cm]{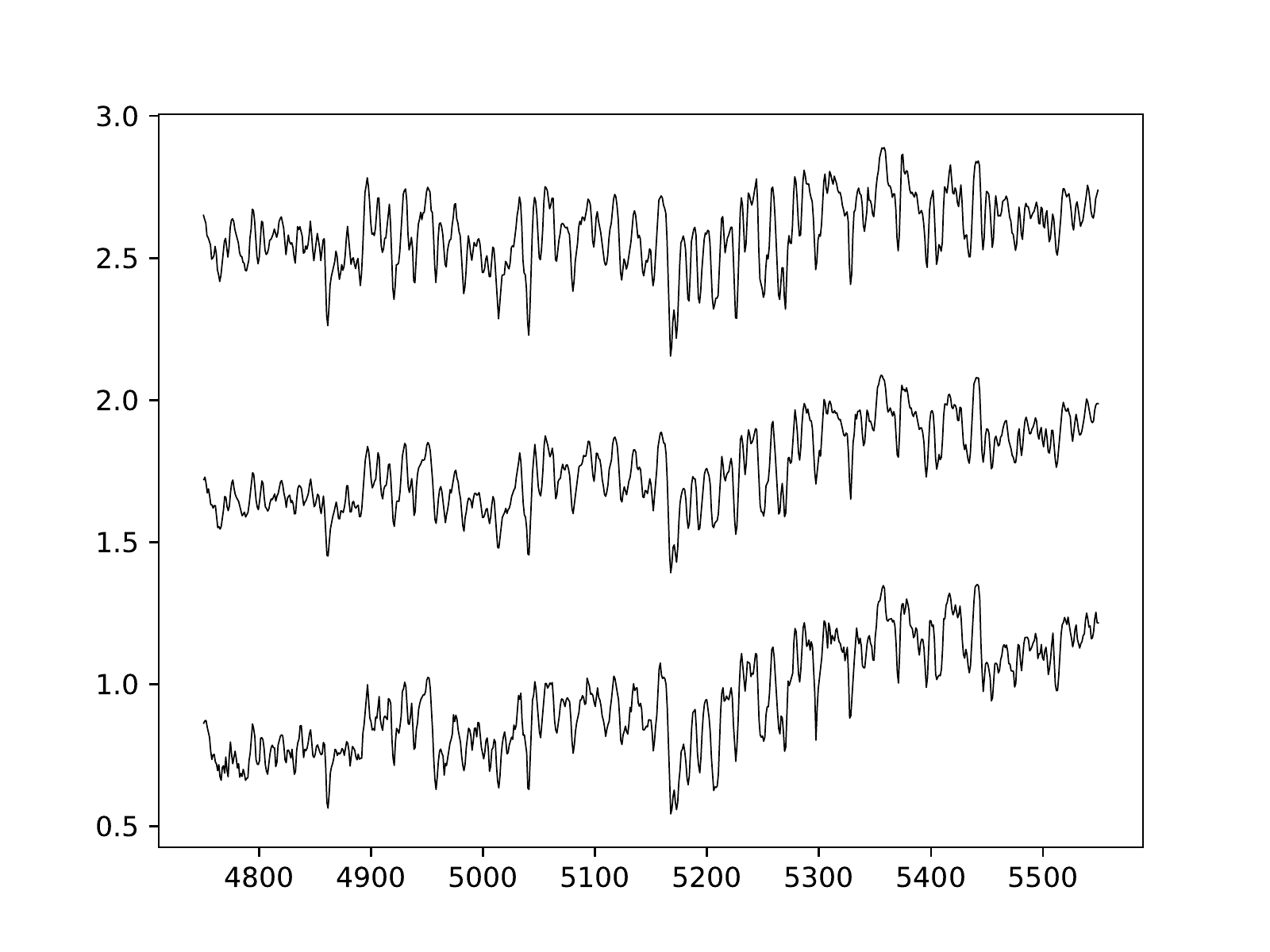}
   \caption{Spectral type sequence for supergiants from the SMC at $\lambda/\delta\lambda\sim1\,300$. The stars are shown  from top to bottom: (upper panel) SMC214 (G5\:Iab) [2012 field-3-smc 334], SMC028 (G7\:Iab\,--\,Ib) [2012 field-3-smc 160], and YSG007 (K0\:Iab\,--\,Ib) [2012 field-3-smc 325]; (lower panel)   SkKM78 (K3\:Iab) [2012 field-2-smc 253], 55275 (K5\:Iab) [2012 field-2-smc 386], and SMC381 (M1\:Iab) [2012 field-2-smc 396]. The wavelength is given in angstrom.}   
   \label{SMC_LR}
\end{figure*}

\begin{figure*}
   \centering
   \includegraphics[trim=1.8cm 0.4cm 1.5cm 1.3cm,clip,width=12.5cm]{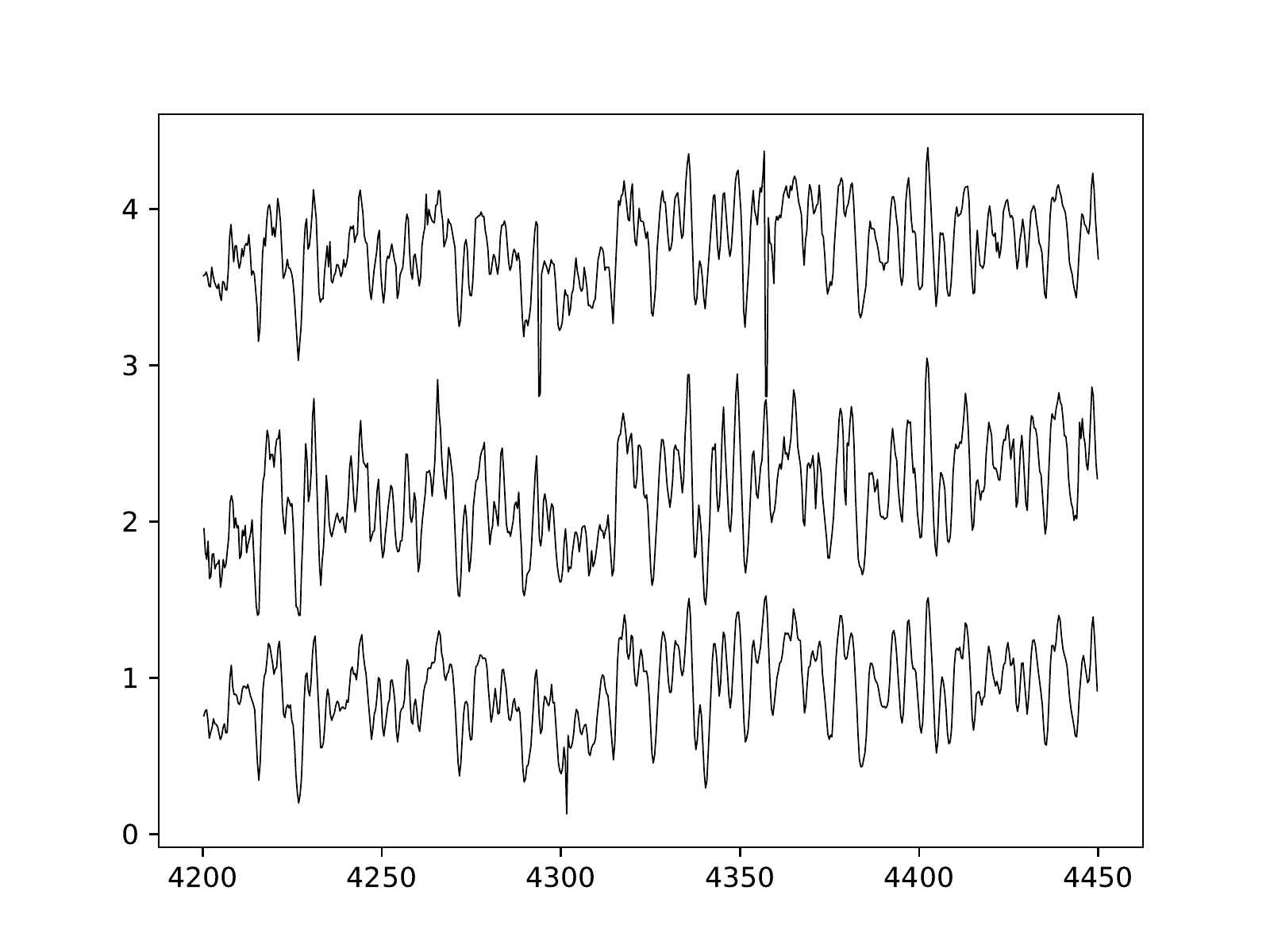}
   \includegraphics[trim=1.8cm 0.4cm 1.5cm 1.3cm,clip,width=12.5cm]{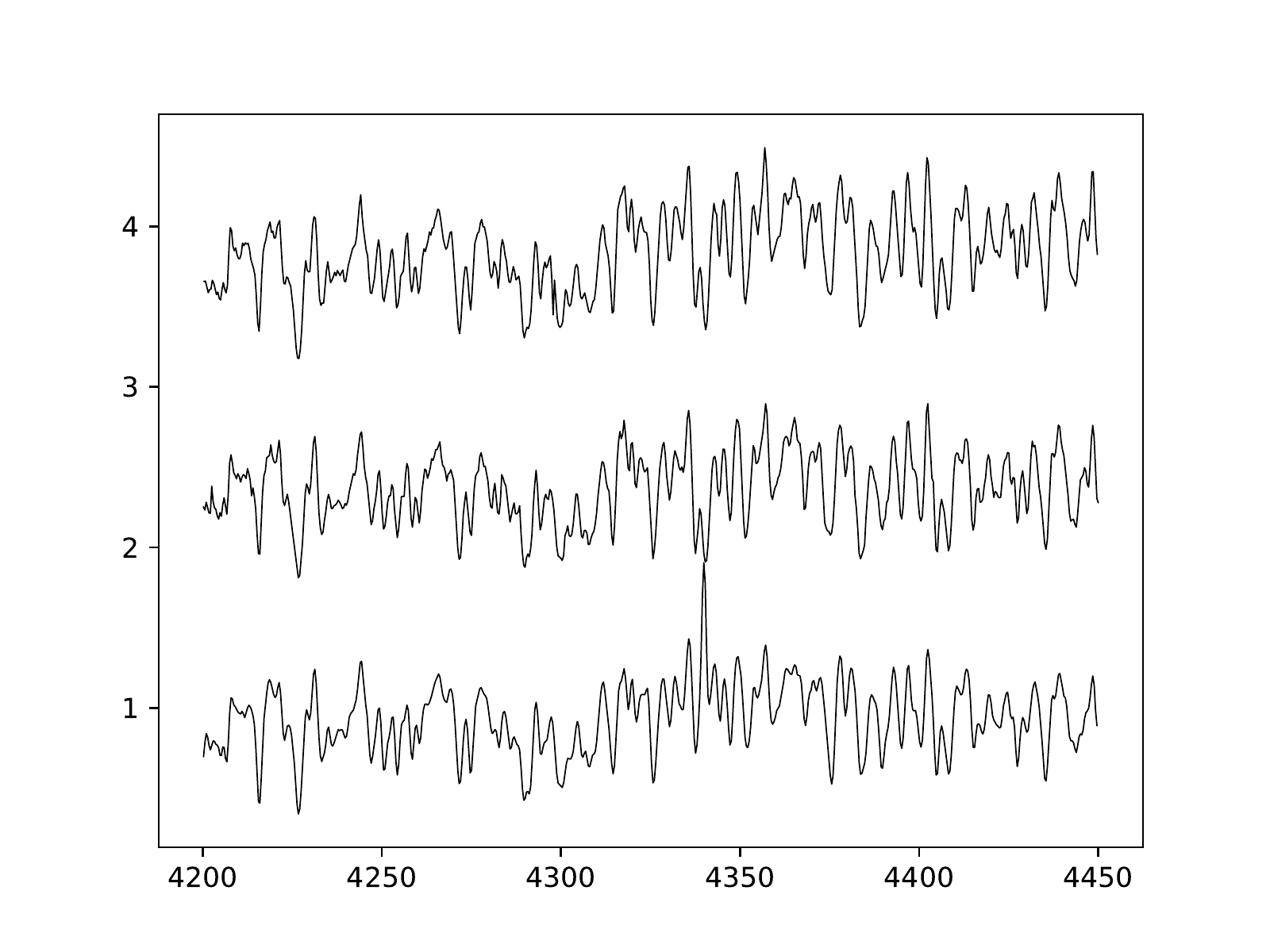}  
   \caption{Spectral type sequence for supergiants from the SMC at $\lambda/\delta\lambda\sim3\,700$ in the range from $4200$ to $4450$~\AA{}. The stars are shown from top to bottom: (upper panel) 61296 (G5\:Iab) [2011 field-1-smc 356], 30135 (G8\:Ia\,--\,Iab) [2012 field-1-smc 115], and 49033 (K1\:Iab) [2012 field-1-smc 325]; (lower panel) 50840 (K3\:Iab) [2012 field-1-smc 340], 11939 (K5\:Iab) [2012 field-1-smc 138], and 18592 (M1\:Iab) [2012 field-1-smc 186] (taking into account that H$\gamma$ is in emission in this star). The wavelength is given in angstrom.}
   \label{SMC_HR_temp}
\end{figure*}

\begin{figure*}
   \centering
   \includegraphics[trim=1.8cm 0.4cm 1.5cm 1.3cm,clip,width=12.5cm]{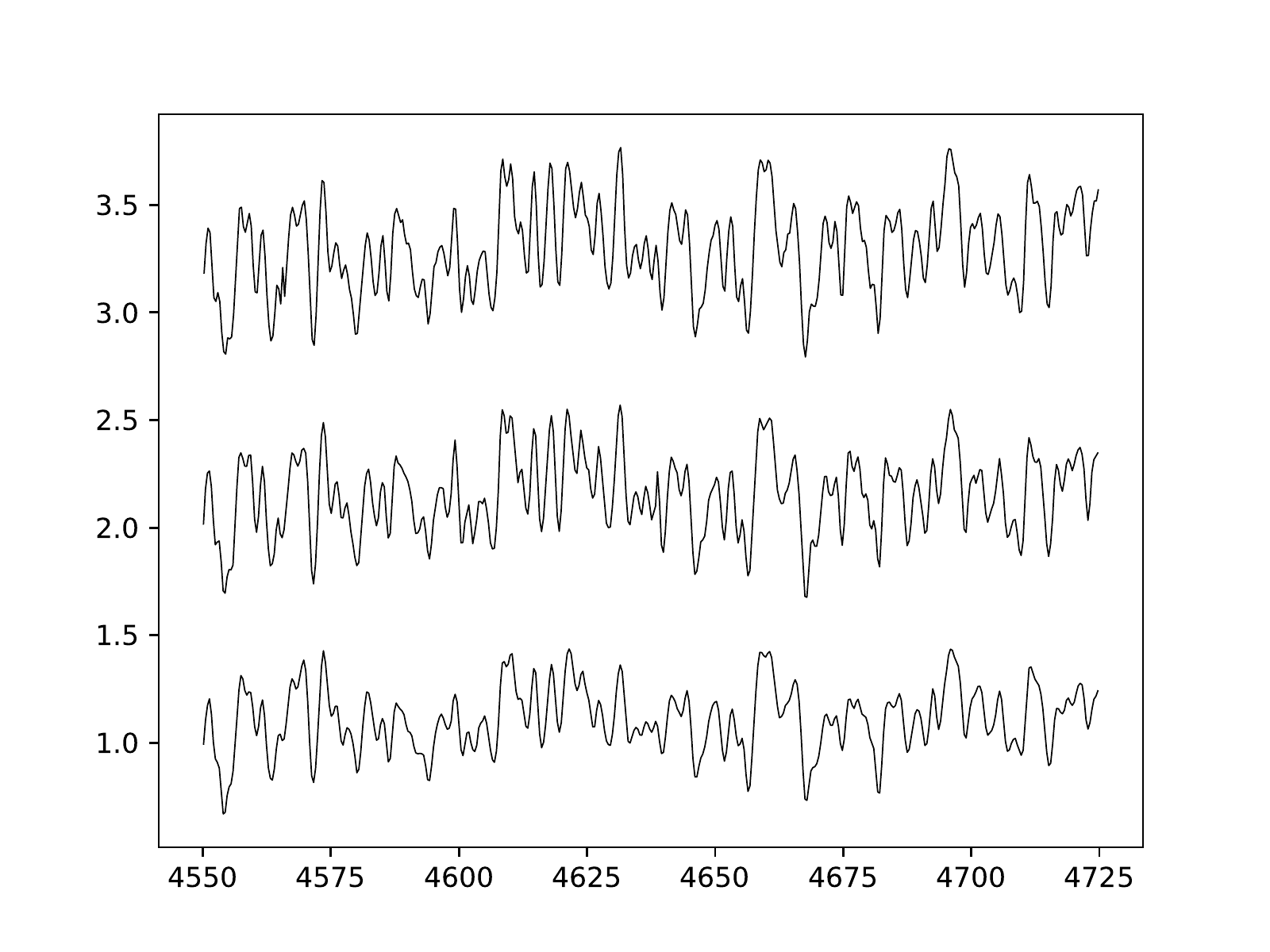}  
   \includegraphics[trim=1.8cm 0.4cm 1.5cm 1.3cm,clip,width=12.5cm]{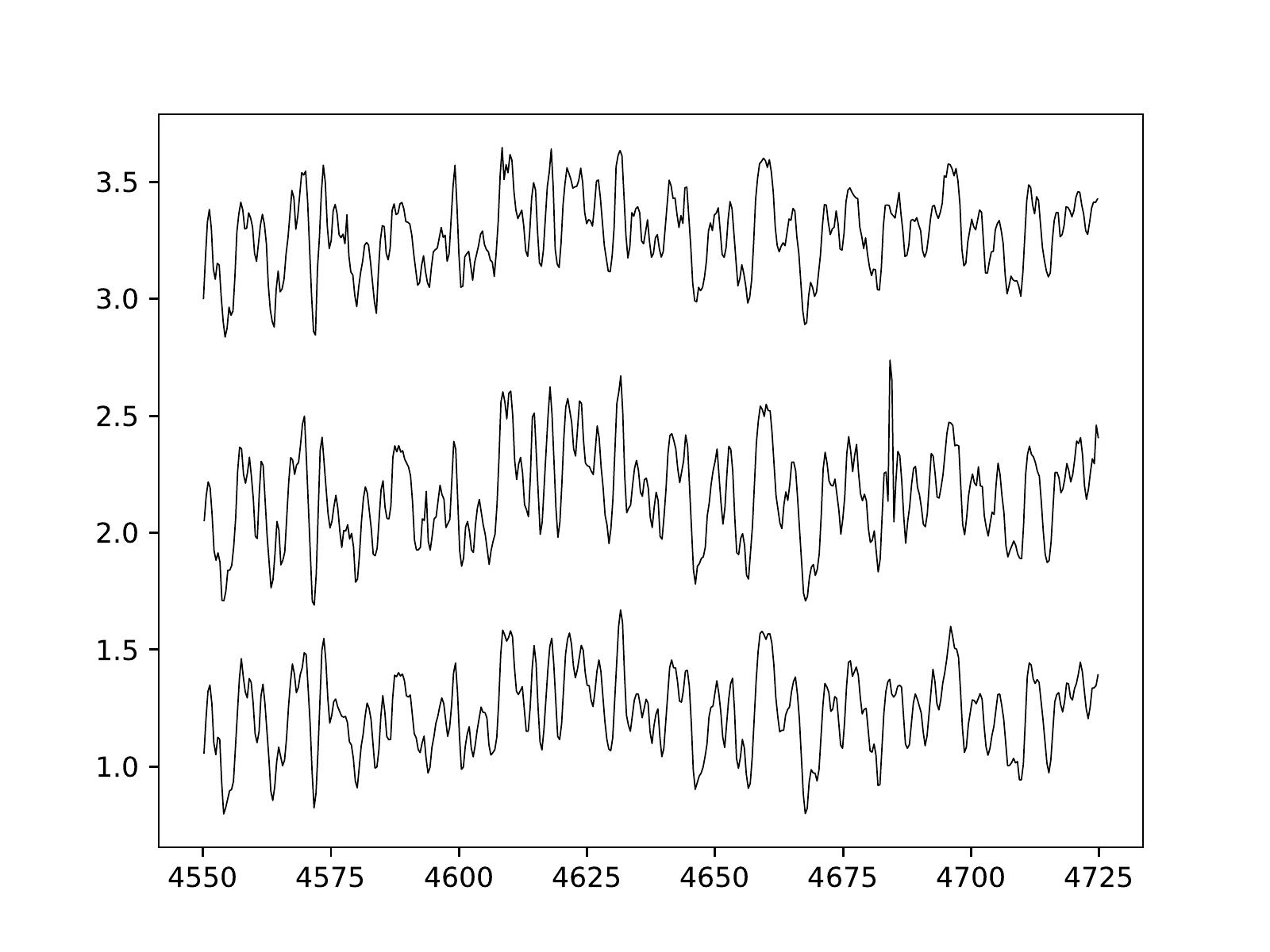}   
   \caption{As in Fig.~\ref{SMC_HR_temp}, but for the range from $4550$ to $4725$~\AA{}}
   \label{SMC_HR_temp}
\end{figure*}

\pagebreak
\section{Spectra in this atlas and their classification}

\begin{longtab}
\begin{landscape}
\begin{longtable}[h]{c c c | c c c | c c c | c c | c c c }
\caption{Sample of the list of the in the atlas and their properties (the full version is available at the CDS). The ''Name in catalogue" indicates the name of that spectrum in \citetalias{gon2015}. The column ''Fibre" indicates the identification number of the fibre assigned to that target in that field. The tag ''C" in SpT and LC columns is for Carbon or S~stars.\label{tab_total}}\\
\hline\hline
\noalign{\smallskip}
\multicolumn{3}{c|}{Spectra identifier}&\multicolumn{3}{c|}{Target identifier}&\multicolumn{3}{c|}{Radial velocities}&\multicolumn{2}{c|}{Classification}&\multicolumn{3}{c}{\textit{Gaia} information}\\
\multicolumn{3}{c|}{}&Name&&&\multicolumn{3}{c|}{(km\,s$^{-1}$)}&Spectral&Luminosity&&\multicolumn{2}{c}{RV$_{\textit{Gaia}}\pm$error}\\
Year&Field&Fibre&in catalogue&RA&DEC&$V_{0}$&$V_{\rm HEL}$&$V_{\rm LSR}$&type&class&Source Id.&\multicolumn{2}{c}{(km\,s$^{-1}$)}\\
\noalign{\smallskip}
\hline
\noalign{\smallskip}
\endfirsthead
\caption{Continuation of Table~\ref{tab_total}}\\
\noalign{\smallskip}
\multicolumn{3}{c|}{Spectra identifier}&\multicolumn{3}{c|}{Target identifier}&\multicolumn{3}{c|}{Radial velocities}&\multicolumn{2}{c|}{Classification}&\multicolumn{3}{c}{\textit{Gaia} information}\\
\multicolumn{3}{c|}{}&Name&&&\multicolumn{3}{c|}{(km\,s$^{-1}$)}&Spectral&Luminosity&&\multicolumn{2}{c}{RV$_{\textit{Gaia}}\pm$error}\\
Year&Field&Fibre&in catalogue&RA&DEC&$V_{0}$&$V_{\rm HEL}$&$V_{\rm LSR}$&type&class&Source Id.&\multicolumn{2}{c}{(km\,s$^{-1}$)}\\
\noalign{\smallskip}
\hline
\noalign{\smallskip}
\endhead
2010&field-1-smc&009&67554&01:08:14.650&-72:46:40.80&204&206&193&K2&Ia\,--\,Iab&4687414400757153664&197.4&1.2\\
2010&field-1-smc&010&58738&01:04:15.460&-72:45:19.90&191&192&179&G4&Ia&4687423471738176512&173&6\\
2010&field-1-smc&013&60447&01:04:53.050&-72:47:48.50&178&179&167&K2&Iab&4687411450138497280&174.4&0.8\\
2010&field-1-smc&018&58472&01:04:09.520&-72:50:15.30&189&190&177&G7&Ia\,--\,Iab&4687410900382741888&189.4&2.4\\
2010&field-1-smc&019&63114&01:06:01.370&-72:52:43.20&201&202&190&K5&Iab&4687411965536012544&194.6&0.9\\
2010&field-1-smc&023&56389&01:03:27.610&-72:52:09.40&154&155&142&K2&Iab&4687409285475098112&151.6&0.8\\
2010&field-1-smc&031&49990&01:00:54.130&-72:51:36.30&195&196&183&G6.5&Ia&4685923669156145536&--&--\\
2010&field-1-smc&032&53557&01:02:23.710&-72:55:21.20&181&182&169&K1&Iab&4687408765770723200&169.9&1.4\\
2010&field-1-smc&042&43219&00:58:23.300&-72:48:40.70&145&146&133&K3&Iab&4685973598219151232&140.9&0.7\\
2010&field-1-smc&114&32188&00:55:03.710&-73:00:36.60&170&171&158&G8&Ia\,--\,Iab&4685961808462981248&157.9&0.5\\
2010&field-1-smc&119&26778&00:53:24.560&-73:18:31.60&153&153&140&G7&Iab&4685887664493672704&153.8&1.0\\
2010&field-1-smc&124&30135&00:54:26.900&-72:52:59.40&158&158&146&G8&Ia\,--\,Iab&4685978198050211584&150.6&0.6\\
2010&field-1-smc&128&25550&00:53:02.850&-73:07:45.90&144&144&132&K3&Iab&4685937413045456384&--&--\\
2010&field-1-smc&129&27443&00:53:36.440&-73:01:34.80&152&152&140&K4&Ia\,--\,Iab&4685942085965189888&--&--\\
2010&field-1-smc&130&25888&00:53:09.040&-73:04:03.60&168&168&156&K4&Ia\,--\,Iab&4685940922099984640&164.0&1.8\\
2010&field-1-smc&131&15510&00:50:06.420&-73:28:11.10&169&170&157&K3&Ia\,--\,Iab&4685878932782750208&--&--\\
2010&field-1-smc&132&20133&00:51:29.680&-73:10:44.30&174&174&162&M1.5&Ia&4685938340755997824&165.4&2.0\\
2010&field-1-smc&133&12572&00:49:05.250&-73:31:07.80&235&235&222&K3&III&4685877287851140608&229.8&0.4\\
2010&field-1-smc&134&13740&00:49:30.340&-73:26:49.90&161&161&149&K2&Iab&4685878524801572224&156.7&0.9\\
2010&field-1-smc&136&11709&00:48:46.320&-73:28:20.70&144&144&132&K4&Iab&4685878215564271360&141.0&0.6\\
2010&field-1-smc&140&11939&00:48:51.830&-73:22:39.30&144&144&132&M2&Ia\,--\,Iab&4685926761539086848&--&--\\
2010&field-1-smc&141&13472&00:49:24.530&-73:18:13.50&141&141&129&G7&Ia\,--\,Iab&4685930167498629760&136&3\\
2010&field-1-smc&143&9766&00:48:01.220&-73:23:37.50&144&145&132&K2&Ia\,--\,Iab&4685926525365285760&137.0&1.1\\
2010&field-1-smc&146&12707&00:49:08.230&-73:14:15.50&166&166&153&K1&Ia\,--\,Iab&4685931537540389632&159.3&2.5\\
2010&field-1-smc&148&13951&00:49:34.420&-73:14:09.90&133&133&120&K2&Ia\,--\,Iab&4685932843199703296&125.6&1.9\\
2010&field-1-smc&149&27945&00:53:45.740&-72:53:38.50&140&141&128&G6.5&Iab&4685955486272265984&139.9&1.5\\
2010&field-1-smc&151&10889&00:48:27.020&-73:12:12.30&145&145&133&K3.5&Ia&4685944323714595072&136&3\\
2010&field-1-smc&155&11101&00:48:31.920&-73:07:44.40&146&146&134&K0&Ia\,--\,Iab&4685947794049430656&136.1&0.4\\
2010&field-1-smc&157&8367&00:47:18.110&-73:10:39.30&139&139&127&K3.5&Ia&4685947308658920448&126&3\\
2010&field-1-smc&159&8324&00:47:16.840&-73:08:08.40&139&139&127&G7&Iab&4685948442516792064&--&--\\
2010&field-1-smc&163&8930&00:47:36.940&-73:04:44.30&132&133&120&M3.5&Iab&4688951075680944640&127.8&1.4\\
2010&field-1-smc&164&12322&00:49:00.320&-72:59:35.70&151&151&139&K4&Ia&4688953201648488064&143&4\\
2010&field-1-smc&171&26402&00:53:17.810&-72:46:06.90&156&156&144&K3&Ia\,--\,Iab&4688962581857059840&153.8&0.3\\
2010&field-1-smc&174&19551&00:51:20.230&-72:49:22.10&150&150&138&K0&Iab&4688961379266217984&--&--\\
\noalign{\smallskip}
\hline
\end{longtable}
\end{landscape}
\end{longtab}

\end{document}